\documentclass[twocolumn]{aastex63}

\usepackage{color, colortbl}
\usepackage{url}
\usepackage{longtable}

\usepackage{xcolor}

\newcommand{\msun}{$M_\odot$}

\newcommand{\xmm} {\textit{XMM-Newton}}
\newcommand{\chandra} {\textit{Chandra}}
\newcommand{\hst} {\textit{HST}}

\newcommand{\cmsq} {cm$^{-2}$}
\newcommand{\nh} {\ensuremath{N_{\rm{H,NGC~4993}}}}

\newcommand{\xspec}{{\sc xspec}}

\newcommand{\countss}{\mbox{\thinspace counts\thinspace s$^{-1}$}}

\newcommand{\ergcms}{\mbox{\thinspace erg\thinspace cm$^{-2}$\thinspace s$^{-1}$}}

\definecolor{lgray}{gray}{0.8}

\submitjournal{ApJ}

\shorttitle{\object{GW170817} Panchromatic Afterglow}
\shortauthors{{Makhathini} et al.}

\begin{document}

\title{The Panchromatic Afterglow of \object{GW170817}: \\The full uniform dataset, modeling, comparison with previous results and implications}

\correspondingauthor{S. Makhathini, K. P. Mooley}
\email{sphemakh@gmail.com, kunal@astro.caltech.edu}


\author[0000-0001-9565-9622]{S.\ Makhathini}
\affil{Department of Physics \& Electronics, Rhodes University, Makhanda, 6139, South Africa}
\affil{South African Radio Astronomy Observatory (SARAO), Cape Town, Observatory, 7925, South Africa}
\affil{School of Physics, University of the Witwatersrand, Johannesburg, Braamfontein, 2000, South Africa}

\author[0000-0002-2557-5180]{K.~P.\ Mooley}
\altaffiliation{Jansky Fellow (NRAO/Caltech).}
\affil{National Radio Astronomy Observatory, Socorro, New Mexico 87801, USA}
\affil{Caltech, 1200 E. California Blvd. MC 249-17, Pasadena, CA 91125, USA}

\author[0000-0002-8147-2602]{M.\ Brightman}
\affil{Caltech, 1200 E. California Blvd. MC 249-17, Pasadena, CA 91125, USA}

\author[0000-0002-2502-3730]{K.\ Hotokezaka}
\affil{Department of Astrophysical Sciences, Princeton University, Peyton Hall, Princeton, NJ 08544, USA}
\affil{Research Center for the Early Universe, Graduate School of Science, University of Tokyo, Bunkyo-ku, Tokyo 113-0033, Japan}

\author[0000-0002-8070-5400]{A.~J.\ Nayana}
\affil{Department of physics, United Arab Emirates University, Al-Ain, UAE, 15551.}
\affil{National Centre for Radio Astrophysics, Tata Institute of Fundamental Research, PO Box 3, Pune, 411007, India.}

\author[0000-0002-5880-2730]{H.~T.\ Intema}
\affil{International Centre for Radio Astronomy Research, Curtin University, GPO Box U1987, Perth WA 6845, Australia}

\author[0000-0003-0699-7019]{D.\ Dobie}
\affil{Sydney Institute for Astronomy, School of Physics, University of Sydney, Sydney, New South Wales 2006, Australia.}
\affil{ATNF, CSIRO Astronomy and Space Science, PO Box 76, Epping, New South Wales 1710, Australia}
\affil{ARC Centre of Excellence for Gravitational Wave Discovery (OzGrav), Hawthorn, Victoria, Australia}

\author[0000-0002-9994-1593]{E.\ Lenc}
\affil{ATNF, CSIRO Astronomy and Space Science, PO Box 76, Epping, New South Wales 1710, Australia}

\author[0000-0001-8472-1996]{D.~A.\ Perley}
\affil{Astrophysics Research Institute, Liverpool John Moores University, IC2, Liverpool Science Park, 146 Brownlow Hill, Liverpool L3 5RF, UK}

\author[0000-0002-4223-103X]{C.~Fremling}
\affiliation{Division of Physics, Mathematics and Astronomy, California Institute of Technology, Pasadena, CA 91125, USA}

\author[0000-0002-8079-7608]{J.\ Moldon}
\affil{Instituto de Astrof\'isica de Andaluc\'ia (IAA, CSIC), Glorieta de las Astronom\'ia, s/n, E-18008 Granada, Spain}
\affil{Jodrell Bank Centre for Astrophysics, School of Physics and Astronomy, University of Manchester, Manchester M13 9PL, UK}

\author[0000-0002-9190-662X]{D.\ Lazzati}
\affil{Department of Physics, Oregon State University, 301
  Weniger Hall, Corvallis, OR 97331, USA}

\author[0000-0001-6295-2881]{D.~L.\ Kaplan}
\affil{Center for Gravitation, Cosmology, and Astrophysics, Department of Physics, University of Wisconsin-Milwaukee, P.O. Box 413, Milwaukee, WI 53201, USA}

\author[0000-0003-0477-7645]{A.\ Balasubramanian}
\affil{Department of Physics and Astronomy, Texas Tech University, Box 1051, Lubbock, TX 79409-1051, USA}

\author[0000-0003-1095-8194]{I.~S.\ Brown}
\affil{Center for Gravitation, Cosmology, and Astrophysics, Department of Physics, University of Wisconsin-Milwaukee, P.O. Box 413, Milwaukee, WI 53201, USA}

\author[0000-0002-6575-4642]{D.\ Carbone}
\affil{University of the Virgin Islands, 2 Brewers Bay Road, Charlotte Amalie, USVI 00802, USA}

\author[0000-0002-0844-6563]{P.\ Chandra}
\affil{National Centre for Radio Astrophysics, Tata Institute of Fundamental Research, Pune University Campus, Ganeshkhind Pune 411007, India}

\author[0000-0001-8104-3536]{A.\ Corsi}
\affil{Department of Physics and Astronomy, Texas Tech University, Box 1051, Lubbock, TX 79409-1051, USA}

\author[0000-0002-1873-3718]{F.\ Camilo}
\affil{South African Radio Astronomy Observatory (SARAO), Cape Town, Observatory, 7925, South Africa}

\author[0000-0001-9434-3837]{A.\ Deller}
\affil{Centre for Astrophysics \& Supercomputing Swinburne University of Technology John St, Hawthorn VIC 3122 Australia}
\affil{ARC Centre of Excellence for Gravitational Wave Discovery (OzGrav), Hawthorn, Victoria, Australia}

\author{D.~A.\ Frail}
\affil{National Radio Astronomy Observatory, Socorro, New Mexico 87801, USA}

\author[0000-0002-2686-438X]{T.\ Murphy}
\affil{Sydney Institute for Astronomy, School of Physics, University of Sydney, Sydney, New South Wales 2006, Australia.}
\affil{ARC Centre of Excellence for Gravitational Wave Discovery (OzGrav), Hawthorn, Victoria, Australia}

\author[0000-0001-7089-7325]{E.~J.\ Murphy}
\affil{National Radio Astronomy Observatory, Charlottesville, VA 22903, USA}

\author[0000-0002-4534-7089]{E.\ Nakar}
\affil{The Raymond and Beverly Sackler School of Physics and Astronomy, Tel Aviv University, Tel Aviv 69978, Israel}

\author[0000-0003-1680-7936]{O.\ Smirnov}
\affil{Department of Physics \& Electronics, Rhodes University, Makhanda, 6139, South Africa}
\affil{South African Radio Astronomy Observatory (SARAO), Cape Town, Observatory, 7925, South Africa}

\author{R.~J.\ Beswick}
\affil{Jodrell Bank Centre for Astrophysics, Department of Physics and Astronomy, The University of Manchester, M13 9PL, UK.}

\author{R.\ Fender}
\affil{Denys Wilkinson Building, Keble Road, Oxford OX1 3RH, UK}
\affil{Department of Astronomy, University of Cape Town, Private Bag X3, Rondebosch 7701, South Africa}

\author[0000-0002-7083-4049]{G.\ Hallinan}
\affil{Caltech, 1200 E. California Blvd. MC 249-17, Pasadena, CA 91125, USA}

\author[0000-0001-6864-5057]{I.\ Heywood}
\affil{Subdepartment of Astrophysics, Denys Wilkinson Building, Keble Road, Oxford OX1 3RH, UK}
\affil{Department of Physics \& Electronics, Rhodes University, Makhanda, 6139, South Africa}
\affil{South African Radio Astronomy Observatory (SARAO), Cape Town, Observatory, 7925, South Africa}

\author[0000-0002-5619-4938]{M.\ Kasliwal}
\affil{Caltech, 1200 E. California Blvd. MC 249-17, Pasadena, CA 91125, USA}

\author[0000-0003-1954-5046]{B.\ Lee}
\affil{Korea Astronomy and Space Science Institute, Daedeokdae-ro 776, Yuseong-gu, Daejeon 34055, Republic of Korea}
\affil{MS314-6, Infrared Processing and Analysis Center, California Institute of Technology, Pasadena, CA 91125, USA}

\author[0000-0002-1568-7461]{W.\ Lu}
\affil{Caltech, 1200 E. California Blvd. MC 249-17, Pasadena, CA 91125, USA}

\author[0000-0001-5605-1809]{J.\ Rana}
\affil{104 Davey Lab, Box C-77, Penn State University, PA 16802, USA}

\author[0000-0002-3623-0938]{S.\ Perkins}
\affil{South African Radio Astronomy Observatory (SARAO), Cape Town, Observatory, 7925, South Africa}

\author[0000-0002-2340-8303]{S.~V.\ White}
\affil{Department of Physics \& Electronics, Rhodes University, Makhanda, 6139, South Africa}

\author[0000-0003-0608-6258]{G.~I.~G.\ J\'ozsa}
\affil{Department of Physics \& Electronics, Rhodes University, Makhanda, 6139, South Africa}
\affil{South African Radio Astronomy Observatory (SARAO), Cape Town, Observatory, 7925, South Africa}
\affil{Argelander-Institut f\"ur Astronomie, Auf dem H\"ugel 71, D-53121 Bonn, Germany}

\author[0000-0002-2933-9134]{B.\ Hugo}
\affil{Department of Physics \& Electronics, Rhodes University, Makhanda, 6139, South Africa}
\affil{South African Radio Astronomy Observatory (SARAO), Cape Town, Observatory, 7925, South Africa}

\author{P.\ Kamphuis}
\affil{Ruhr-Universit\"at Bochum, Faculty of Physics and Astronomy, Astronomical Institute, 44780 Bochum, Germany}





\begin{abstract}
We present the full panchromatic afterglow light curve data of GW170817, including new radio data as well as archival optical and X-ray data, between 0.5 and 940 days post-merger. 
By compiling all archival data, and reprocessing a subset of it, we have evaluated the impact of differences in data processing or flux determination methods used by different groups, and attempted to mitigate these differences to provide a more uniform dataset.
Simple power-law fits to the uniform afterglow light curve indicate a $t^{0.86\pm0.04}$ rise, a $t^{-1.92\pm0.12}$ decline, and a peak occurring at $155\pm4$ days.
The afterglow is optically thin throughout its evolution, consistent with a single spectral index ($-0.584\pm0.002$) across all epochs.
This gives a precise and updated estimate of the electron power-law index, $p=2.168\pm0.004$.
By studying the diffuse X-ray emission from the host galaxy, we place a conservative upper limit on the hot ionized ISM density, $<$0.01\,cm$^{-3}$, consistent with previous afterglow studies.
Using the late-time afterglow data we rule out any long-lived neutron star remnant having magnetic field strength between 10$^{10.4}$~G and 10$^{16}$~G.
Our fits to the afterglow data using an analytical model that includes VLBI proper motion from \citeauthor{mooley2018-vlbi}, and a structured jet model that ignores the proper motion, indicates that the proper motion measurement needs to be considered while seeking an accurate estimate of the viewing angle.
\end{abstract}


\section{Introduction} \label{sec:intro}
Discovered on August 17, 2017 and localized to the lenticular galaxy \object{NGC 4993} at 40~Mpc \citep{coulter2017}, \object{GW170817} is the first binary neutron star merger detected in gravitational waves \citep{abbott2017a}.
Uniquely, \object{GW170817} was also accompanied by radiation across the electromagnetic spectrum \citep{abbott2017e}, which allowed the merger astrophysics to be studied in great detail.
A low-luminosity short $\gamma$-ray burst \cite[SGRB;][]{goldstein2017,savchenko2017} was observed 1.7 seconds after the merger.
The macronova/kilonova, which peaked at ultraviolet (infrared) wavelengths on timescales of a few hours (days), indicated $\sim$0.05 \msun~of r-process enriched merger ejecta traveling at 0.1$c$--0.3$c$ \citep[e.g.][]{arcavi2017, cowperthwaite2017, drout2017, kasen2017, kasliwal2017, nicholl2017, pian2017, smartt2017, soares-santos2017, tanvir2017, valenti2017, villar2017}.

The synchrotron afterglow, first detected 9 days after the merger at X-ray wavelengths \citep{troja2017}, 16 days post-merger in the radio \citep{hallinan2017}, and 110 days post-merger in the optical \citep{lyman2018}, gave key insights into the relativistic ejecta and the circum-merger environment.
The delayed onset and rising light curve of the afterglow ruled out an on-axis (typical) SGRB jet \citep{hallinan2017,evans2017,troja2017,alexander2017,margutti2017,haggard2017,kim2017,murguia-berthier2017,kim2017,ruan2018,resmi2018,lyman2018,lazzati2018}.
Radio monitoring over the first 100 days after merger ruled out a simple (top-hat) off-axis jet and established that the panchromatic afterglow emission, as well as the $\gamma$-rays, were produced in a mildly relativistic wide-angle outflow \citep{mooley2018-wideoutflow}.
Such an outflow could be explained by a cocoon \citep[e.g.,][]{lazzati2017,gottlieb2018} formed due to the interaction between an ultrarelativistic jet (as seen in SGRBs) and the merger dynamical/wind ejecta or due to the fast tail of the (fairly isotropic) dynamical ejecta.
The afterglow emission peaked and started to decline approximately 160 days post-merger \citep{ddobie2018, alexander2018, davanzo2018, nynka2018, troja2018}.
While the steeply declining light curve disfavored the isotropic ejecta model, the light curve and polarization measurements remained inconclusive as to whether a putative jet successfully penetrated the merger ejecta or was completely choked by it \citep[e.g.,][]{margutti2018, alexander2018, corsi2018, nakarpiran2018, lamb2018}.

The degeneracy between the successful- and choked-jet models \citep[e.g.][]{nakar2018, gill_granot2018} was finally broken through the measurement of superluminal motion, at four times the speed of light between 75--230 d post-merger, of the radio source using Very Long Baseline Interferometry (VLBI) \citep{mooley2018-vlbi}.
The light curve and VLBI modeling \citep{mooley2018-vlbi} together indicated that the jet core was successful and narrow, having an opening angle $<5$\,degrees and observed from a viewing angle between 14--28\,degrees, and energetic, with the isotropic equivalent energy being about $10^{52}$\,erg \citep[lying at the tail-end of the regular SGRB distribution;][]{fong2015}.
The implied Lorentz factor close to the light curve peak is $\Gamma\simeq4$ \citep{mooley2018-vlbi}.
The strong constraints on the geometry of \object{GW170817} facilitated a precise measurement of the Hubble constant \citep{hotokezaka2019}.
Subsequently, independent VLBI and afterglow light curve observations \citep{ghirlanda2019,mooley2018-strongjet,troja2019,lamb2019,fong2019,hajela2019} confirmed the presence of a successful jet in the late-time afterglow of \object{GW170817}.

The wealth of observational data collected for the afterglow of \object{GW170817} makes this one of the best studied (off-axis) SGRB afterglows.
However, the dataset currently available in literature lacks uniformity, i.e. it suffers from differences in data processing and flux determination methods used by different groups. 
Recently, \cite{fong2019} and \cite{hajela2019} presented the reprocessing of some of the optical and X-ray afterglow data ({\it Hubble Space Telescope} or {\it HST} F606W 600\,nm data and {\it Chandra X-ray Observatory} soft X-ray data), but the majority of the data (including radio data) were still lacking uniformity.
Further, various groups have modeled the afterglow data of \object{GW170817}, but these groups have used different subsets of the data.
The impact of these inhomogeneities is seen (at least partially) in the significant differences in the modeling results \citep[e.g.,][see below]{resmi2018,mooley2018-vlbi,ghirlanda2019,wu2019,lamb2019}.
Taken together, a thorough compilation of all the observational data and a uniform dataset for the afterglow of \object{GW170817} is warranted.

In this work we present a thorough compilation of the available radio, X-ray and optical data. The work includes new data not published before, and a reprocessing of some previous data sets using consistent methodology. The result is a fairly uniform panchromatic dataset of \object{GW170817}'s afterglow. 
The observational data span 0.5 days to 940 days post-merger.
We have made these afterglow measurements available in ASCII format on the web\footnote{\url{https://github.com/kmooley/GW170817/} or \url{http://www.tauceti.caltech.edu/kunal/GW170817/}}, and this online dataset will be continuously updated (beyond 940 days) as new measurements become available.
The new observations, data compilation and (re)processing are presented in \S\ref{sec:data}.
The full uniform afterglow data are presented in Table~\ref{tab:alldata} and the light curve is shown in Figure~\ref{fig:lc}.
\S\ref{sec:modeling} describes power-law fits to the afterglow light curve and an analytical model to obtain jet and interstellar medium (ISM) parameters.
Constraints on the density of the circum-merger environment and the nature of the merger remnant are presented in \S\ref{sec:obs-constraints}.
In \S\ref{sec:literature} we present preliminary fits to the afterglow light curve using the numerical structured jet model from \cite{lazzati2018}, a short review of all previous modeling efforts, and an examination of  our modeling results in the context of  previous results.
We end with a summary and discussion in \S\ref{sec:discussion}.





\section{Data Compilation, (Re)Processing, Analysis} \label{sec:data}

We compiled all flux density upper limits from the literature (see references given in Table~\ref{tab:alldata}).
Flux densities in the case of radio afterglow detections were compiled from \cite{mooley2018-strongjet} and references therein, and optical (\hst/F606W) afterglow detections reported in the \cite{fong2019} reprocessing.
Below we report on new data obtained with the Karl G.\ Jansky Very Large Array (VLA), MeerKAT, the Australia Telescope Compact Array (ATCA) and \textit{enhanced} Multi Element Remotely Linked Interferometer Network (eMERLIN) radio telescopes between 180 and 780 days post-merger spanning frequencies between 1.2--9\,GHz.

We further reprocessed and analyzed radio data reported by  \cite{resmi2018,margutti2018,alexander2018}, ensuring consistent method of flux determination \cite[as reported in][]{mooley2018-strongjet}.
Similarly, we also reprocessed X-ray and optical (\hst/F814W) data to ensure a uniform data processing and flux determination technique.
Our radio reprocessing substantially improves the precision of the flux density values (by up to a factor of 2 in RMS noise, i.e., 1$\sigma$ errorbar) with respect to previously-published values.
Through our reprocessing, we find discrepancies of up to 1.5$\sigma$ in the previously-published radio flux density values.
Our measurements with the reprocessed X-ray and optical data are in agreement, within 1$\sigma$, with previously published values.

The new observations and data (re)processing methods are described below, and the full afterglow dataset spanning radio, optical and X-ray frequencies is given in Table~\ref{tab:alldata}.
We note that, in Table~\ref{tab:alldata}, all flux density measurements are quoted with 1$\sigma$ error bars and all upper limits are 3$\sigma$.
The full uniform afterglow light curve is shown in Figure~\ref{fig:lc}. 

\begin{table*}
    \centering
    \begin{tabular}{ccccccccc}
    \hline
      RA & Dec & $F_{0.7}$ & $F_{1.3}$ & $F_{3}$ & $F_{6}$ & $ F_{7.2}$ & $ F_{10}$ & $F_{15}$ \\\hline
      13h09m53.9s & 23d21m34s & $530\pm10$ & $910\pm20$ & $565\pm15$ & $280\pm10$ & $300\pm30$ & $160\pm5$ & $70\pm 10$\\
      13h09m44.5s & 23d24m09s & $440\pm20$ & $230\pm15$ & $120\pm10$ &  $65\pm5$ & $70\pm10$ & $60\pm5$ & $37\pm5$ \\\hline
    \end{tabular}
    \caption{Positions, peak flux densities (mean and standard deviations; $\mu$Jy/beam) at 0.7, 1.3, 3, 6, 7.2, 10 and 15~GHz for the two reference sources used to bootstrap the fluxscale of the radio data (VLA, MeerKAT and uGMRT) processed in this work.}
    \label{tab:bootstrap}
\end{table*}

\subsection{VLA} \label{sec:data:vla}
VLA data of \object{GW170817} covering the period between 2017 August 18 and 2018 Jan 8 have been reported by \cite{alexander2017, hallinan2017, mooley2018-wideoutflow, mooley2018-vlbi, margutti2018, alexander2018, mooley2018-strongjet}; see Table \ref{tab:alldata}. 
We have reprocessed some of these observations (see Table \ref{tab:alldata}) using the NRAO Common Astronomy Software Applications ({\tt CASA}) pipeline \citep{mcmullin2007} (version 5.4) and {\tt WSClean} \citep{offringa2014} for imaging\footnote{Although we used {\tt WSClean}, we noted that unresolved background radio sources in the field did not vary significantly with respect to the images generated using {\tt CASA} {\tt clean} (which was used for all other VLA data), thus ensuring uniformity of flux measurements for all VLA data. In order to quantify the imaging differences between the two software, we reimaged a few calibrated datasets with both {\tt WSClean} and {\tt CASA} {\tt clean}, keeping the imaging parameters similar to the ones we used for the other reprocessed datasets. We found the mean flux density difference of sources within the FWHM of the primary beam to be $<$2\%.}



Additionally, we observed \object{GW170817} on 2018 Dec 18--20 and 2019 Sep 24--27 with the VLA (PI: Corsi; VLA/18B-204). 
The Wideband Interferometric Digital Architecture (WIDAR) correlator was used at S band (2--4\,GHz). We used \object[PKS J1248-1959]{PKS~J1248$-$1959} as the phase calibrator and \object{3C286} as the flux density and bandpass calibrator. The data were calibrated and flagged for radio frequency interference (RFI) using the {\tt CASA} pipeline (version 5.4). We then split and imaged the target data using the {\tt CASA} tasks {\tt split} and {\tt clean}.

For all datasets, imaging involved Briggs weighting with a robust value between 0--0.5, and a threshold of 3x the thermal noise.
For any radio image (VLA and other telescopes, described below), we measured the peak flux density of a point source (e.g. GW170817 and comparison sources) as the pixel value at the actual source position, as appropriate for point sources.
The associated uncertainty is the RMS noise in a source-free region of the image in the vicinity of the target.
Note that VLA absolute flux density calibration is accurate to about 5\% at L-band through to Ku-band \citep[1--18 GHz]{perley2017}.

\subsection{ATCA} \label{sec:data:atca}

We observed \object{GW170817} with the ATCA (PI: Dobie, Piro) at four epochs between 2018 Nov to 2019 Sep (Table~\ref{tab:alldata}). 
We determined the flux scale and bandpass response for all epochs using the ATCA primary calibrator \object[PKS B1934-638]{PKS~B1934$-$638}. Observations of \object[PKS B1245-197]{PKS~B1245$-$197} were used to calibrate the complex gains. All observations used two bands of 2048\,MHz centered at 5.5 and 9.0\,GHz. 

We reduced the visibility data using standard {\tt MIRIAD} \citep{sault1995} routines. 
The calibrated visibility data from both bands were combined, averaged to 32\,MHz channels, and imported into DIFMAP \citep{shepherd1997}. 
Bright field sources were modeled separately for each band using the visibility data and a combination of point-source and Gaussian components with power-law spectra. After subtracting the modeled field sources from the visibility data, \object{GW170817} dominates the residual image. Restored naturally-weighted images for each band were generated by convolving the restoring beam and modeled components, adding the residual map and averaging to form a wide-band image. Image-based Gaussian fitting with unconstrained flux density and source position was performed in the region near \object{GW170817}. Note that the absolute flux density measurements from ATCA are accurate to about 5\% \citep{partridge2016}.

Following our Markov chain Monte Carlo (MCMC) analysis \citep[\S\ref{sec:modeling}; see also][]{mooley2018-strongjet} we corrected all ATCA flux density values with a constant multiplicative factor of 0.8. 

\subsection{uGMRT} \label{sec:data:gmrt}

We reprocessed archival upgraded Giant Metrewave Radio Telescope (uGMRT) Band 5 (1.0--1.4\,GHz) data
with the {\tt CASA} package. The data were initially flagged and calibrated using a custom developed pipeline in {\tt CASA}\footnote{\url{http://www.ncra.tifr.res.in/~ishwar/pipeline.html}}. The data were further inspected for RFI and flagged using standard tasks in {\tt CASA}.
The target source data were then imaged with the {\tt CASA} task {\tt clean}. A few rounds of phase-only self-calibration and two rounds of amplitude and phase self-calibration were done in order to approach thermal noise. The flux density of the GW source at multiple epochs is listed in Table \ref{tab:alldata}. 

 The uGMRT Band 4 (0.55--0.85\,GHz) observations were processed using the SPAM pipeline \citep{intema2009,intema2017} by splitting the wideband data in 6~frequency chunks of 50\,MHz wide which are processed separately. For each observing session, instrumental calibrations are derived using the best available scan on flux calibrator \object{3C147} or \object{3C286}. These calibrations are applied to the target visibility data, after which this data is split into separate files. The target files per epoch are concatenated and taken through several cycles of self-calibration, imaging, and flagging of bad data. The final two cycles include direction-dependent calibration to mitigate ionospheric effects. The pipeline yields an image and calibrated visibility dataset for each frequency chunk. In a final step, to benefit from the improved sensitivity and $uv$-coverage of the wideband data, the calibrated visibility data of the 6~frequency chunks per epoch are jointly imaged using WSClean \citep{offringa2014}. Note that the uGMRT measurements have a systematic uncertainty between 10-15\% \citep[][in this work we assume 10\% for model fits]{chandra2017}.


\subsection{MeerKAT} \label{sec:data:meerkat}
\object{GW170817} was observed with the MeerKAT telescope \citep{jonas2018, camilo2018} at 7 epochs between 2018 Jan 18 and 2018 Sep 02 (see Table \ref{tab:alldata}). The first observation was performed during the AR1 phase using 16 antennas, while remaining observations used the full 64 antenna array. All observations are centred at 1.3\,GHz using 4096 channels spanning 856\,MHz and an 8\,s integration time. About 10\% of the band is flagged due to the bandpass roll off, resulting in an effective bandwidth of 770MHz, and a further 27\% of is flagged due to RFI. At 1.3\,GHz, the field of view (full width at half maximum of the primary beam) is about 1.1 degrees.  The data are processed using the {\it Containerized Automated Radio Astronomy Calibration} pipeline  \citep[{\tt CARACal};][]{ramatsoku2020}, which performs; i) automatic RFI flagging using {\tt CASA} and {\tt AOFlagger} \citep{offriga2010}; ii) a standard cross-calibration (delay, bandpass and gain calibration) using a combination of {\tt CASA} and {\tt MeqTrees} \citep{noordam2010}. We used \object[PKS 1934-638]{PKS~1934$-$638} as the primary calibrator and \object{3C286} as the secondary calibrator; and iii) a direction-dependent self-calibration \citep{pearson1984} that uses a combination of {\tt WSClean} \citep{offringa2014}, {\tt CubiCal} \citep{kenyon2018}, and {\tt PyBDSF} \citep{mohan2015}. After the cross-calibration step, we found a variability of around 10\% on the flux density measurements between epochs, which was corrected by bootstrapping the fluxes to a common fluxscale using 2 reference point-like sources (see Table \ref{tab:bootstrap}) within 2 arcminutes of the afterglow position. It is worth noting that this uncertainty is due to our calibration process, and is not a limitation of the telescope. The self-calibration includes a model of the MeerKAT primary beam that is derived from Holography measurements of the array \citep{asad2021}.

\subsection{eMERLIN} \label{sec:data:emerlin}

 We observed \object{GW170817} with the eMERLIN array between January and March 2018 with 11 individual runs. Each run had a duration of 5--6\,hours. Observations were conducted using the the C band receiver tuned at frequencies between 4.82 and 5.33\,GHz, for a total bandwidth of 512\,MHz distributed in 4 spectral windows, each one divided into 512 channels. The phase reference source was \object[J1311-2329]{J1311$-$2329}. Flux density calibration and bandpass correction were obtained from \object{3C286} and \object{OQ208}, respectively. The observations were primarily at low elevations ($<20$\,degrees), and the flux density measurements may be affected by a small bias due to reduced gain sensitivity of the telescopes at these elevations. Nevertheless, the core of the host galaxy was detected (at $160\pm20\,\mu$Jy\,beam$^{-1}$) in almost all runs, with associated variability of about 12\% between runs, compatible with the expected uncertainties \citep{garrington2004,muxlow2020}.
We measure the flux density of \object{NGC 4993} to be $0.25\pm0.01\,$mJy\,beam$^{-1}$ at 4.5\,GHz with the VLA \citep{mooley2018-vlbi}, indicating a flux density correction factor of about 0.6 for the eMERLIN measurements.
The flux density of GW170817 (detected only in the first observing run, on 2018 January 14), the associated uncertainty, and 3$\sigma$ upper limits reported in Table~\ref{tab:alldata} include the absolute flux density error (25\%) and statistical map noise error.

\begin{figure*}
\centering
\includegraphics[width=\textwidth]{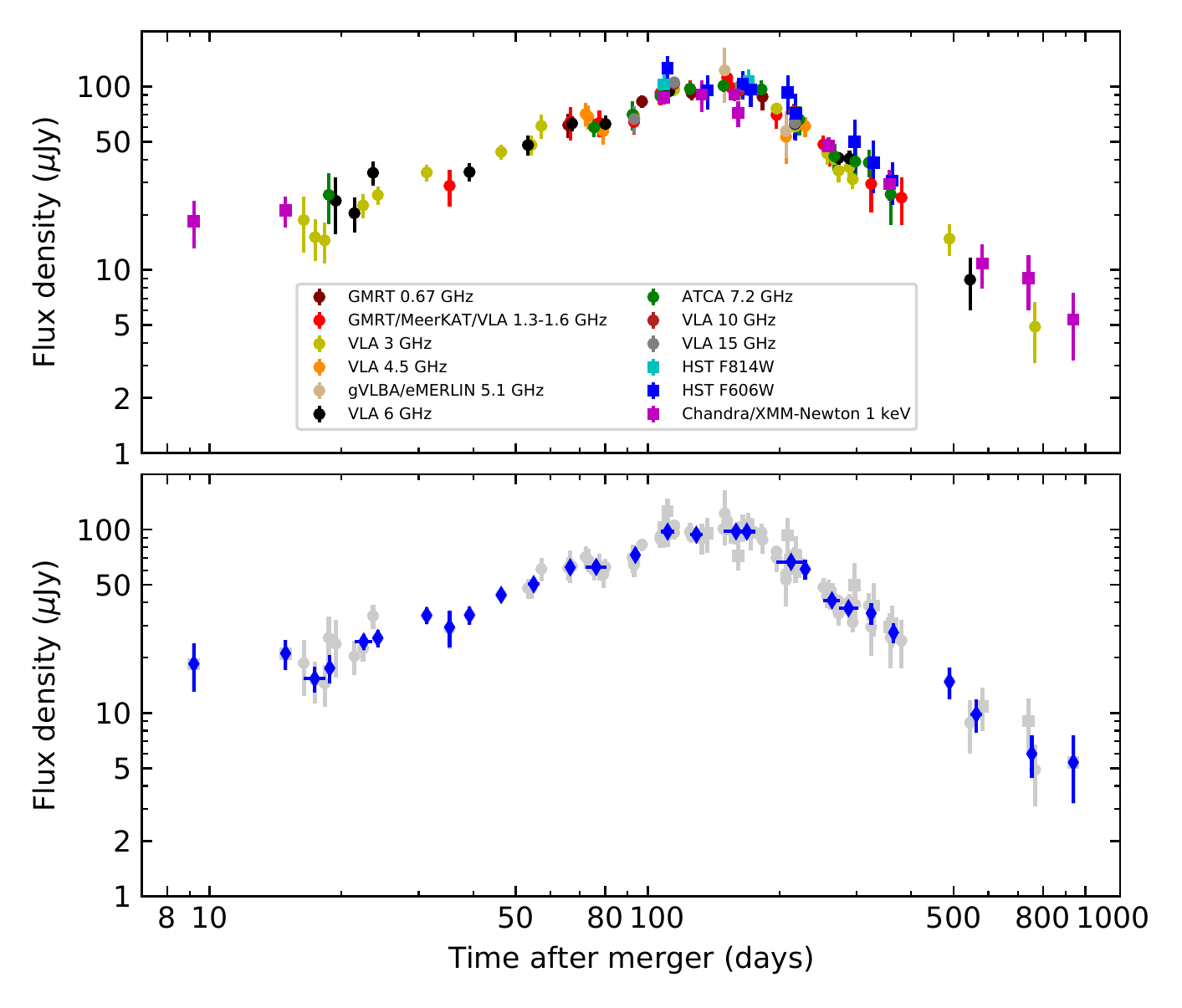}\\
\caption{{\it Upper panel}: The panchromatic (radio, optical and X-ray) afterglow light curve of \object{GW170817}, color coded according to the observing frequency, up to 940 days post-merger (all data points have 1$\sigma$ errorbars as presented in Table~\ref{tab:alldata}; upper limits are not shown here) using the uniform dataset presented in this work. The light curve is scaled to 3~GHz using the best-fit spectral index ($-0.584$) derived from the MCMC power-law fitting (see \S\ref{sec:modeling}).
{\it Lower panel}: The averaged (using moving average; $\Delta t/t=1/15$ where $\Delta t$ is the width of the kernel and $t$ is the time after merger) light curve (blue data points) shows a general trend consistent with power-law rise and decline. 
In grey are the same data points as shown in the upper panel.}
\label{fig:lc}
\end{figure*}

\subsection{HST} \label{sec:data:hst}
 
Reduced HST images were downloaded from the MAST archive.  To remove most of the stellar light we first fit a simple Sersic model to \object{NGC 4993} using Galfit \citep{peng2002}.  This leaves significant residuals (asymmetries, dust lanes, shell/tidal features, etc.) which were removed by applying a 1\arcsec~box median filter.  This was followed by astrometric correction to align the images with each other. 
To obtain PSF photometry at the expected position of \object{GW170817} we have done the following: (1) we estimate an empirical PSF model ($50\times50$\,pixel size) by first detecting point-like sources using SExtractor \citep{bertin1996}, and constructing an average PSF from these. Sources that do not fit well to the average PSF were removed, and a final PSF was constructed using the remaining point-like sources. (2) we fit the PSF model to the data at the expected position of \object{GW170817}. Uncertainties and limiting magnitudes are estimated by randomly ($>$ 100 times) fitting the PSF model to the background as close to the position of \object{GW170817} as possible.\footnote{We have validated our PSF photometry method by comparing the resulting lightcurve of \object{GW170817} in the F606W filter with that published by \cite{fong2019}. The methods agree within the uncertainties.  For F606W we preferred to use the \citeauthor{fong2019} flux density values for our final panchromatic dataset (Table~\ref{tab:alldata}) due to more precise background subtraction (availability of an observation template for the subtraction of the host galaxy, as done by \citeauthor{fong2019}). Nevertheless, the precision of the flux density measurements is comparable for both methods ($\sim$20\%).}

In Table~\ref{tab:alldata} we report the upper limit from a previously-unpublished dataset. 
The observations (PI: N. Tanvir) were carried out with the WFC3/UVIS detector using the F814W on 2018 August 08.4 and have a total exposure time of 5.2 ks.

\subsection{Chandra and XMM-Newton} \label{sec:data:x-ray}
We list the \chandra\ and \xmm\ observational data on \object{GW170817} used here for spectral analysis in Table \ref{tab:table_xray_obsdata}. 


For \chandra\ we used {\sc ciao} v4.13 \citep{2006SPIE.6270E..1VF} with {\sc caldb} v4.9.5 to analyze the data, initially reprocessing the data using the {\tt chandra\_repro} tool. We then astrometrically aligned the events of each individual obsID to a common frame, which is important when the X-ray emission from GW170817 is not detected in an individual observation. To do this we first ran {\tt wavdetect} on events in the 0.5--8 keV range with wavelet scales of 1, 2 and 4 pixels and all other parameters set as default to get the positions of the X-ray sources detected in each observation. This yielded $\sim$200--500 source positions depending on the exposure. We then used {\tt wcs\_match} to obtain the transform matrix, and the source list from obsID 20860 as the reference, as done in Hajela et al (2021), filtering to sources within 1\arcsec\ of each other and a residual limit of 1\arcsec. The typical residual was then $\sim$0.5\arcsec. We used {\tt wcs\_update} and the resulting transform matrix to align the astrometry of each obsID to the reference frame.

The tool {\tt specextract} was used to extract the X-ray spectra of  \object{GW170817} and its host galaxy \object{NGC 4993}. For \object{GW170817}, we use a circular region with radius $1\arcsec$, which encompasses 90\% of the PSF at 1.5\,keV, centered on the source. We extract background events from a nearby source-free circular region with radius $19.2\arcsec$. We do not weight the ARFs generated by {\tt specextract} (`weight=no'), which is appropriate for a point-like source. In an earlier version of this paper, we set `psfcorr=yes', however this was determined to overestimate the PSF correction\footnote{https://cxc.harvard.edu/ciao/ahelp/specextract.html}, which is likely the reason for the discrepancy in fluxes found by Troja et al (2021). We therefore used the tool {\tt arfcorr} as a workaround to correct the ARFs and incorporate the PSF correction.

For groups of observations made close to each other, in order to increase signal to noise, we combine the spectral products using the {\sc ciao} tool {\tt combine\_spectra} for use in spectral fitting. 


To measure the emission from \object{NGC 4993}, we extract events from a $50\arcsec$ circular region centered on the nucleus, masking 6 point sources lying within the region (including \object{GW170817} and the AGN associated with the galaxy). We used a circular source-free region, with radius $34\arcsec$, outside the galaxy extraction region, to estimate the background. For the emission from the galaxy, since we do not expect this to change significantly over time, we combine spectral products from all \chandra\ observations.


We used {\sc xmmsas} v18.0.0 to analyze the \xmm\ data and the tool {\tt evselect} to extract spectral and lightcurve data. We do not use \xmm\ obsID 0830191001 since the source is too faint with respect to the Active Galactic Nucles (AGN) for spectral analysis. Count rates greater than 0.7\,s$^{-1}$ in the range 10--12\,keV on the pn detector were used to determine periods of high background data, which were excluded from our analysis. Significant background flaring resulted in only 26 and 48\,ks filtered exposures for obsIDs 0811210101 and 0811212701, respectively. Source events were then extracted from a circular region with radius 5$\arcsec$ centered on \object{GW170817}. While this is less than 50\% of the EPIC-pn encircled energy, it was necessary to use a small region in order to exclude emission from the AGN, which is only 10~arcsec from the source. In order to account for the AGN, we extracted background events from a region at the same distance from the AGN as GW170817. 

For both \object{GW170817} and \object{NGC 4993}, the spectra were grouped with a minimum of 1 count per bin with the {\sc heasoft} tool {\tt grppha}. We used the X-ray spectral fitting package {\sc xspec} v12.10.1 to fit the data. For \object{GW170817} we fit the data with an absorbed power-law model ({\tt tbabs*ztbabs*powerlaw}), where {\tt tbabs} is a neutral absorbing column attributed to our own Galaxy, fixed at $7.59\times10^{20}$\,\cmsq\ (HI4PI Collaboration, N. Ben Bekhti, L. Floer, et al., 2016, Astronomy \& Astrophysics, 594, A116 - HI4PI Map), and {\tt ztbabs} is a neutral absorbing column intrinsic to the source, \nh, at $z=9.73\times10^{-3}$ which we initially allow to vary within the fit. We use the Cash statistic as the fit statistic, with the background subtracted, and the spectra were fitted in the 0.5--8\,keV range for \chandra\ and the 0.2--10\,keV range for \xmm. 

We initially allow all spectral parameters to vary across observational epochs to test for spectral variations, however we do not find any evidence for this. We therefore re-fit the spectra with the \nh\ and $\Gamma$ parameters tied across all epochs. We find no evidence for absorption intrinsic to the source with a 90\% upper limit of $\nh<5\times10^{20}$\,\cmsq. For the power-law, the best fit photon index is $\Gamma=1.62^{+0.13}_{-0.09}$ (1-$\sigma$ errors) which is consistent with $\Gamma=1.584$ that is inferred from the radio to X-ray spectrum (confirmed in \S\ref{sec:modeling}). Therefore we fix\footnote{We believe that $\Gamma=1.584$ is robust, but to understand the effect of changing $\Gamma$ we did a test. We recalculated the X-ray flux density values by changing $\Gamma$ by 0.2 (i.e. we used $\Gamma=1.38$ and $\Gamma=1.78$) and then redid the MCMC fitting of the full afterglow light curve as described in \S\ref{sec:modeling}. We found that the value of spectral index $\beta$ (reported in Table~\ref{tab:best-fit}) changed by $<1.5\sigma$ and the change in the other fit parameters was negligible, $\ll1\sigma$. We also repeated the MCMC analysis by leaving $\Gamma$ ($=1-\beta$) as a free parameter and found the best-fit value of beta to be the similar ($-0.583\pm0.003$)}. \nh\ (intrinsic)=0 and $\Gamma=1.584$ in order to measure the flux from \object{GW170817}. We calculated the 0.3--10 keV flux and its uncertainty, corrected for the absorption from our own Galaxy, using the model {\tt cflux} in {\sc xspec} which is presented in Table \ref{tab:table_xray_obsdata} along with the observed count rates (i.e. not corrected for PSF losses).

The flux density was determined from the normalization of the power law, which is a measure of the flux density at 1 keV in $\mu$Jy when the commands '{\tt xset pow\_emin 1.0}' and '{\tt xset pow\_emax 1.0}' are used in {\sc xspec}. We used the “error” command in \xspec\ to determine the uncertainties, with a delta-C-stat of 1.0, which corresponds to a 1$\sigma$ confidence level for one free parameter.


The X-ray emission from GW170817 was detected by Chandra in all but four observations, obsIDs 18955, 23183, 24923, and 24924. We calculate the 3$\sigma$ upper limit on the count rate of \object{GW170817} in these observations from events extracted in the background region. We determine that the background count rate in the source extraction region is 6--8$\times10^{-6}$\,\countss. By using the Poisson probability distribution, the 3$\sigma$ upper limit on the source count rate calculated to be 1--2$\times10^{-4}$\,\countss, which when assuming our spectral model corresponds to a 0.3--10\,keV unabsorbed flux of 3--6$\times10^{-15}$\,\ergcms. We present the individual upper limits on the count rates and fluxes in Table \ref{tab:table_xray_obsdata}. As noted in Troja et al (2021), these upper limits are systematically lower than they calculate due to the differing statistical treatment, as they use a Bayesian method and we use classical Poisson statistics.


Overall, we detect 9604 counts between 0.5--8\,keV in the source extraction region from all observations combined of which 9.4\% are attributable to NGC~4993. We use a power-law model to calculate a flux, where the power-law index, $\Gamma=0.7\pm0.6$, and normalization, $N=1.8\pm1.0\times10^{-6}$, yielding a 0.5--8 keV flux of $2.0^{+1.1}_{-0.7}\times10^{-14}$\,erg\,cm$^{-2}$\,s$^{-1}$. Since the source counts are a small fraction of the total (source+background) counts, background subtraction introduces large uncertainties into the spectral modelling results, which should be treated with caution.

We also point the reader to \cite{hajela2019,hajela2020}, who did a similar and independent analysis using the Chandra data. Additionally, we note that the X-ray data have been processed independently also by \cite{troja2017,troja2018,troja2019,piro2019,troja2020-paper,troja2021}. 

\begin{figure*}
\centering
\includegraphics[width=\textwidth]{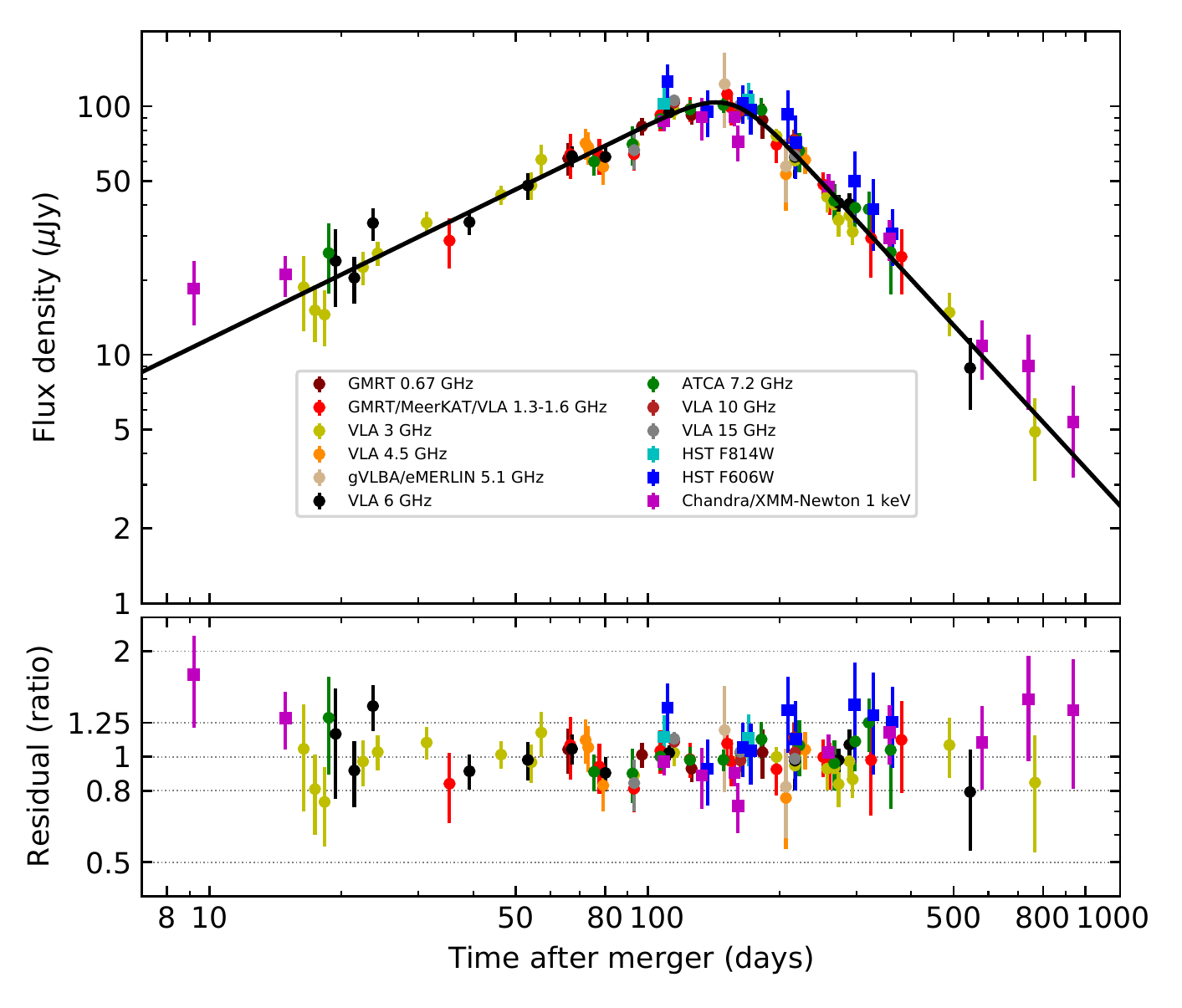}\\
\caption{Broken power-law fit (using MCMC; best-fit shown as a black curve in the upper panel and residual shown in the lower panel) to the afterglow light curve. 
The corner plot for the MCMC analysis is shown in Figure~\ref{fig:lc_bpl_corner}.
The light curve is scaled to 3~GHz using the best-fit spectral index ($-0.584$) derived from the MCMC analysis.
Color coding is the same as in Figure~\ref{fig:lc}. 
The light curve rises as $t^{0.86\pm0.04}$ and declines as $t^{-1.92\pm0.12}$. The light curve peak occurs at $155\pm4$ days post-merger.
The lack of any substantial outlier data points indicates that the afterglow is optically thin throughout its evolution, with the synchrotron self-absorption frequency lying below the radio band and cooling frequency lying above the soft X-ray band.
See \S\ref{sec:modeling} for details.}
\label{fig:lc_bpl}
\end{figure*}

\begin{figure*}
\centering
\includegraphics[width=\textwidth,height=6.5in]{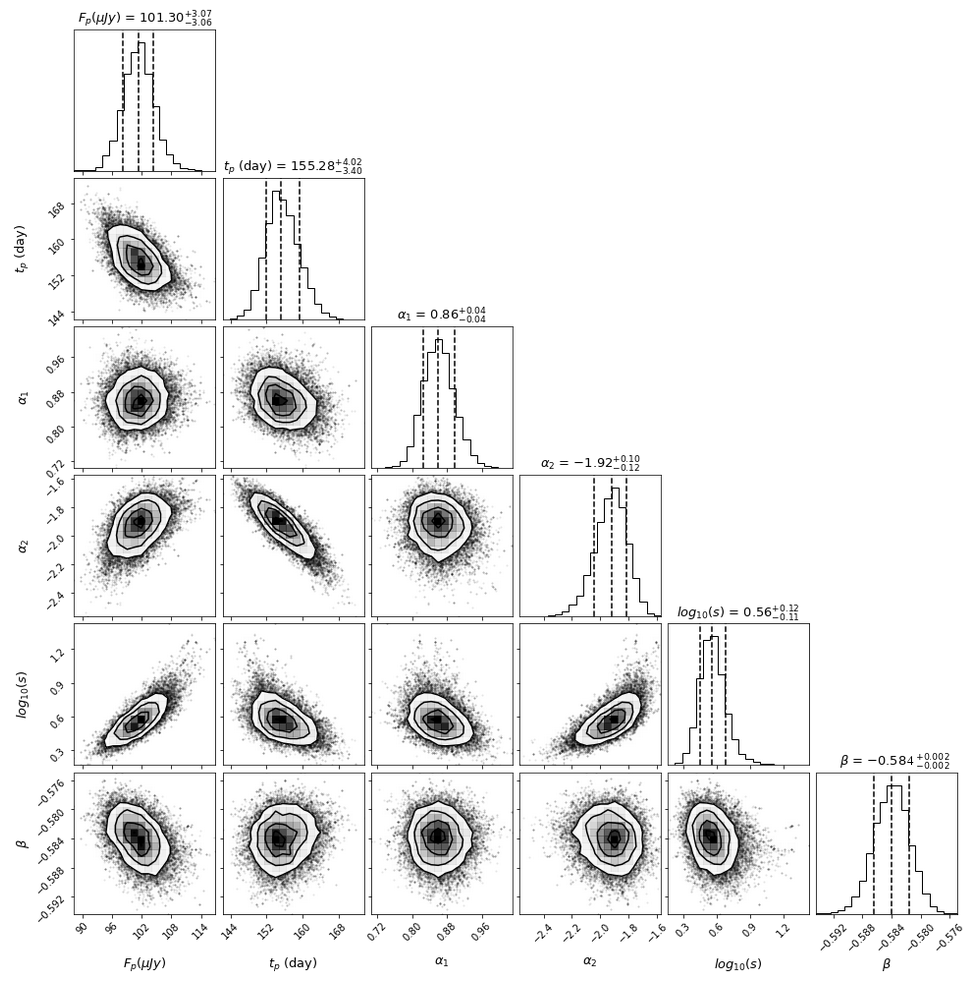}\\
\caption{Corner plot for broken power-law fit to the light curve presented in Figure~\ref{fig:lc_bpl}.  Here, $\beta$ is the spectral index, $F_p$ is the flux density at 3\,GHz at light curve peak, $t_p$ is the  light curve peak time, and $\alpha_1$ and $\alpha_2$ are the power-law rise and decay slopes, respectively.}
\label{fig:lc_bpl_corner}
\end{figure*}

\section{Analytical Modeling} \label{sec:modeling}

Following previous afterglow studies \citep{ddobie2018,alexander2018,mooley2018-strongjet}, we fit the afterglow data\footnote{We do not consider the gVLBA and eMERLIN data points while modeling since they have relatively large uncertainties in the absolute flux calibration.} using a smoothly-broken power law model,

\begin{equation}
    F(t,\nu) = 2^{1/s}\left(\frac{\nu}{3\,{\rm GHz}}\right)^{\beta}F_p~\left[\left(\frac{t}{t_p}\right)^{-s\alpha_1} +\left(\frac{t}{t_p}\right)^{-s\alpha_2}\right]^{-1/s}
\end{equation}  

\noindent where $\nu$ is the observing frequency, $\beta$ is the spectral index, $F_p$ is the flux density at 3~GHz at light curve peak, $t$ is the time post merger, $t_p$ is the  light curve peak time, $s$ is the smoothness parameter, and $\alpha_1$ and $\alpha_2$ are the power-law rise and decay slopes, respectively. This  MCMC fitting was done\footnote{We chose 100 walkers, 1000 steps and flat priors on all of the parameters. Since compact interferometric data may be affected by extended emission from the host galaxy, we also introduced a scale factor into the MCMC fit to explore possible offsets in the ATCA, MeerKAT and uGMRT flux densities. We recover the constant flux multiplication factor of 0.8 for the ATCA \citep[most likely due to the compact array being sensitive to extended structure around GW170817][]{mooley2018-strongjet}, while the factor is consistent with unity for the MeerKAT and uGMRT data. The inclusion of scale factors in the MCMC fitting gave best-fit values and uncertainties of all other parameters in the fit to be almost the same as those reported in Table~\ref{tab:best-fit}, except for parameters $F_{\rm p}$, $t_{\rm p}$ and $log_{10}(s)$ where the {\it uncertainties} were larger by factors of $\sim$2 than the ones given in Table~\ref{tab:best-fit}.} using the Python package {\tt emcee} \citep{foreman-mackey2013}. We obtain best-fit values listed in Table~\ref{tab:best-fit}. 
Figure~\ref{fig:lc_bpl} shows the best-fit broken power-law curve and the resulting residuals. 
Figure~\ref{fig:lc_bpl_corner} shows the corner plot corresponding to the MCMC analysis.

We use a rough analytic model, described in \cite{mooley2018-strongjet}, to estimate the jet opening angle $\theta_j$ and the viewing angle $\theta_v$. The sharpness of the light curve peak, $\Delta t / t = (t_2-t_1)/t_2$, where $t_1$ is the time around the transition from the $t^{0.86}$ rise to the peak and $t_2$ is the time when the light curve approaches $t^{-2}$, is directly related to the ratio $\theta_j$/$\theta_v$.
Using the approximations $\theta_v-\theta_j \gg \theta_j$ and jet Lorentz factor $\Gamma \propto t^{-3/8}$ \citep{blandford1976} we find that $\Delta t / t \simeq (8/3) \theta_j/\theta_v$. Here we use the approximation that $\theta_j/\theta_v$ is much smaller than unity. 
From our MCMC analysis we find $0.2 \lesssim \Delta t / t \lesssim 0.4$ (68\% confidence or better, depending on where $t_1$ and $t_2$ lie), indicating that $0.1 \theta_v \lesssim \theta_j \lesssim 0.2 \theta_v $.
Using $\Gamma\simeq4.1\pm0.5\simeq1/(\theta_v-\theta_j)$ close to the peak of the light curve from the VLBI measurement \citep{mooley2018-vlbi}, we get $\theta_j\simeq 1$--$4\degr$ and $\theta_v\simeq 14$--$20\degr$. 

Using the Blandford-McKee solution \citep{blandford1976}, we can estimate the ratio of the jet kinetic energy and the density of the circum-merger environment $E/n_{\rm ISM}$. We have,

\begin{equation}
E/n_{\rm ISM} \simeq \theta_j^2 R^3 \Gamma^2 m_p c^2 \simeq 8 \theta_j^2 t^3 \Gamma^8 m_p c^5
\end{equation}  

\noindent since $R=\beta c t_{\rm lab} =(1-1/2\Gamma^2) c (1-\beta {\rm cos}(\theta_v))^{-1} t \simeq 2\Gamma^2 c t$, where $R$ is the distance travelled (in the lab frame) by the blastwave, $t_{\rm lab}$ is time in the lab frame, $m_p$ is the proton mass and $c$ is the speed of light.
Hence, $E/n_{\rm ISM} \simeq 1.5 \times 10^{53}$ ($\theta_j/3\degr$)$^2$\,erg\,cm$^3$



\section{Constraints on the merger environment and merger remnant}\label{sec:obs-constraints}

\subsection{ISM density estimate using the diffuse X-ray emission from NGC 4993}

In Section \ref{sec:data:x-ray} we described the X-ray data analysis, including the diffuse X-ray emission from NGC 4993, where we calculate a flux of $2.0^{+1.1}_{-0.7}\times10^{-14}$\,erg\,cm$^{-2}$\,s$^{-1}$, which corresponds to a luminosity of $L_{0.5-8}\sim 2\times 10^{39}\,{\rm erg}\,{\rm s}^{-1}$.

\begin{figure}
\centering
\includegraphics[width=3.5in]{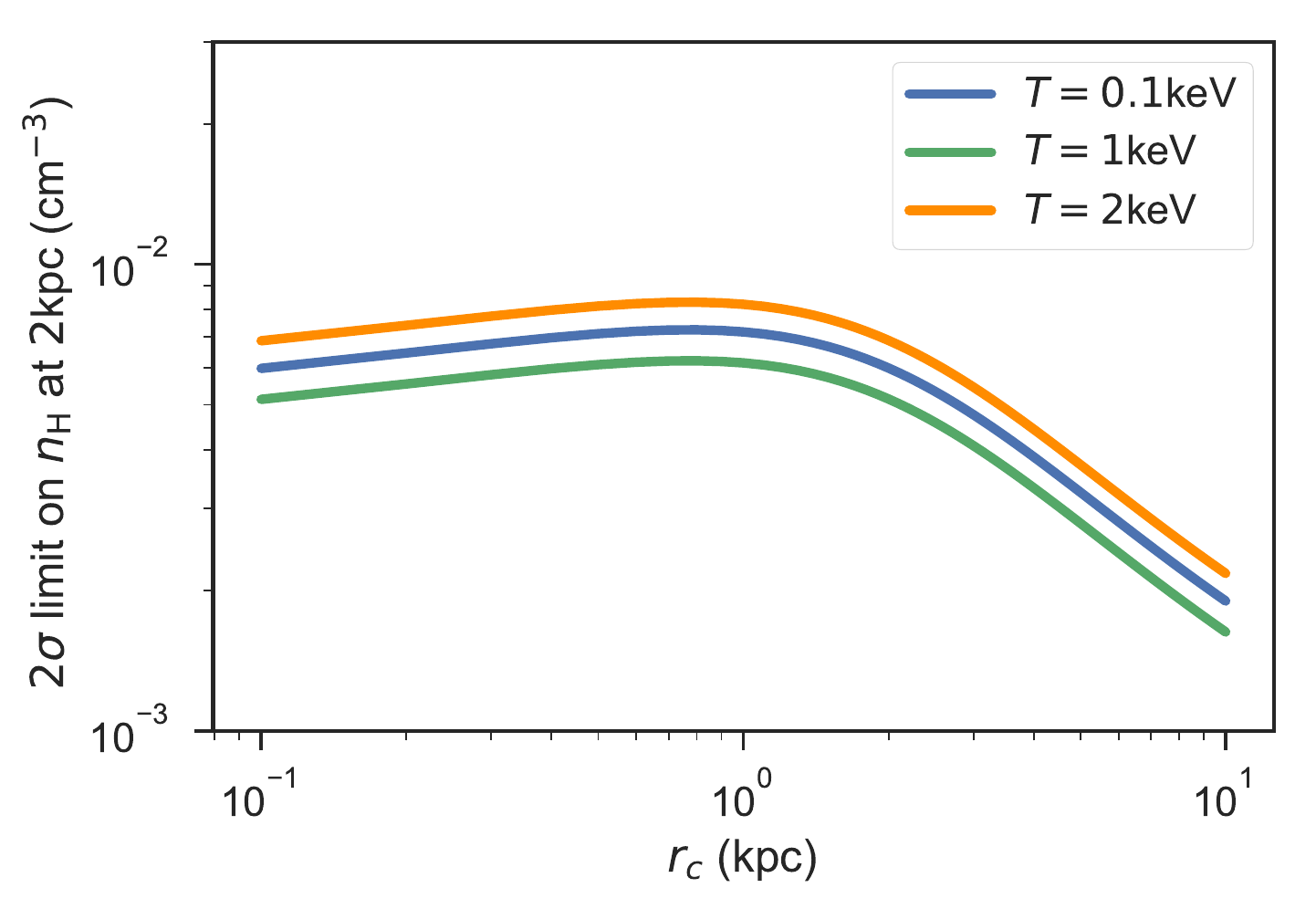}\\
\caption{Upper limits on the number density of hot ionized ISM at the merger location ($\sim 2$~kpc from the center of \object{NGC 4993}) as a function of the core radius ($r_c$) at the temperature ($T$) of the hot ISM.  Here we assume the solar metallicity and $\beta_{\rm ISO}=0.5$.}
\label{fig:density}
\end{figure}

Because of the large uncertainty due to the background subtraction, we cannot separate the diffuse emission of the hot ionized ISM from that arising from unresolved point sources based on their spectral shapes. Therefore, we use the $2\sigma$ upper limit on the X-ray flux  to estimate an limit on the number density of the hot ionized ISM in \object{NGC 4993}.
To obtain the density at the location of the merger from the diffuse emission in the circular region with radius $34\arcsec$, we use an isothermal beta model \citep{Cavaliere1978} to describe the global structure of the hot ionized gas. This model is characterized by the density at the galactic center, the core radius, $r_c$,  the power-law index of the density profile, $\beta_{\rm ISO}$, and temperature $T$.
  For early-type galaxies with a diffuse X-ray luminosity of $\lesssim 10^{40}$\,erg\,s$^{-1}$, the model parameters are typically in the range of $0.1\lesssim T \lesssim 1$\,keV, $r_c\lesssim 10$\,kpc, and $0.25\lesssim \beta_{\rm ISO}\lesssim 1$ \citep{Babyk2018ApJ}. Given the density structure and the X-ray emissivity calculated with {\tt APEC}, we compute the total X-ray flux arising from the hot ionized ISM in the circular region.  Assuming $\beta_{\rm ISO}=0.5$ and the solar metallicity, we obtain a $2\sigma$ upper limit on the ISM density  at the merger, $\lesssim 10^{-2}\,{\rm cm^{-3}}$, as shown in Figure \ref{fig:density}. 
Note that our method provides a good estimate of the mean density of the hot-ionized  ISM but it does not necessarily provide a precise estimate of the density at the merger location.  However, our estimate is  consistent with the result of \cite{hajela2019} --- $n\leq 9.6\times 10^{-3}\,{\rm cm}^{-3}$  --- in which they analyze the X-ray data in a small annular region of the inner part of \object{NGC 4993}. The consistency of these two independent analyses supports  $n\lesssim 10^{-2}\,{\rm cm^{-3}}$. Furthermore,
this estimate is consistent with that estimated from the afterglow light curve and superluminal motion modelings (e.g., \citealt{mooley2018-vlbi}) and from searches for neutral gas ($n<0.04\,{\rm cm}^{-3}$; \citealt{hallinan2017}).

\subsection{The merger remnant}
While GW170817 is believed to have collapsed to a black hole after a short-lived hypermassive neutron star phase, the exact nature of the remnant remains observationally unknown \citep[e.g.][]{kasen2017,yu2018,pooley2018}.
Here we derive constraints on the magnetic field strength ($B$) of any long-lived neutron star remnant\footnote{Here we do not consider any energy loss from gravitational wave radiation. For an alternative analysis which considers such radiation, see \cite{piro2019}.} that may have resulted from \object{GW170817}.
The afterglow flux density measurements up to 940 days post-merger indicate that the late-time afterglow is consistent with a decelerating jet ($t^{-p}$ decline).
The full uniform afterglow light curve is shown in Figure~\ref{fig:lc}.
We consider an upper limit (measured flux density $+$ 2$\sigma$ uncertainty) of 3.6$\times$10$^{-10}$ Jy on the X-ray flux density, corresponding to a luminosity of 2$\times$10$^{38}$\,erg\,s$^{-1}$ for any pulsar wind component.
Pulsar (magnetic dipole) spin-down luminosity is given by \citep{spitkovsky2006,hotokezaka2017,metzger2017},

\begin{equation}
 L_{\rm sd} \simeq 2.3\times10^{43}\,{\rm erg\,s}^{-1} \left(\frac{B}{10^{12}~{\rm G}}\right)^2 \left(\frac{P}{1~{\rm ms}}\right)^{-4} \left(1+\frac{t}{t_{\rm sd}}\right)^{-2}
\end{equation}

where,

\begin{equation}
 t_{\rm sd} \simeq 31~{\rm yr} \left(\frac{B}{10^{12}~{\rm G}}\right)^{-2} \left(\frac{P}{1~{\rm ms}}\right)^{2}
\end{equation}

Assuming the spin period at the time of merger is $P\simeq1$\,ms and that a fraction $f\simeq10^{-2}$ of the spin-down power is converted into X-ray radiation, we obtain\footnote{This lower limit on the magnetic field strength is sensitive to $f$ and $P$. If $f=10^{-3}$ or $P=2$ ms then we get $B\lesssim10^{11}$\,G.}, $B\lesssim10^{10.4}$\,G.

The sensitive late-time afterglow measurement is therefore more constraining than the previous constraints \citep{pooley2018,margutti2018} on the magnetic field strength.
The upper limit of 10$^{10.4}$\,G is at odds with simulations\footnote{We assume here that the small-scale magnetic fields, found in simulations, are comparable in strength to the global dipolar field.} \citep{zrake2013,kiuchi2014,giacomazzo2015} that predict $B\sim10^{15}$--$10^{16}$\,G.
If the simulations accurately represent the dipole magnetic field strength, then it is likely that the merger remnant in \object{GW170817} is a black hole. 

However, we note that we are not able to rule out magnetars with $B\gtrsim10^{16}$\,G since for such objects the spin down luminosity would decrease rapidly within the first $\sim$100\,days (this is a very conservative timescale over which the merger ejecta would become optically thin towards any emission arising from spin-down), below the afterglow luminosity measured for GW170817 (see Figure 7 of \citealt{margutti2018} for example).

 \section{Numerical Modeling}\label{sec:literature}
We have presented the uniform afterglow light curve to the astronomical community with the hope that these data will be used for extensive modeling in the future.
In this section we provide a preliminary\footnote{The model currently excludes VLBI constraints, and assumes constant jet opening angle and blast-wave energy.} update on the jet and ISM parameters obtained using the numerical model from \cite{lazzati2018} (\S\ref{sec:literature:sj}).
In \S\ref{sec:literature:review} we review and contrast with previous modeling efforts.

\subsection{Structured Jet Model}\label{sec:literature:sj}
\begin{figure}
\centering
\includegraphics[width=3.5in]{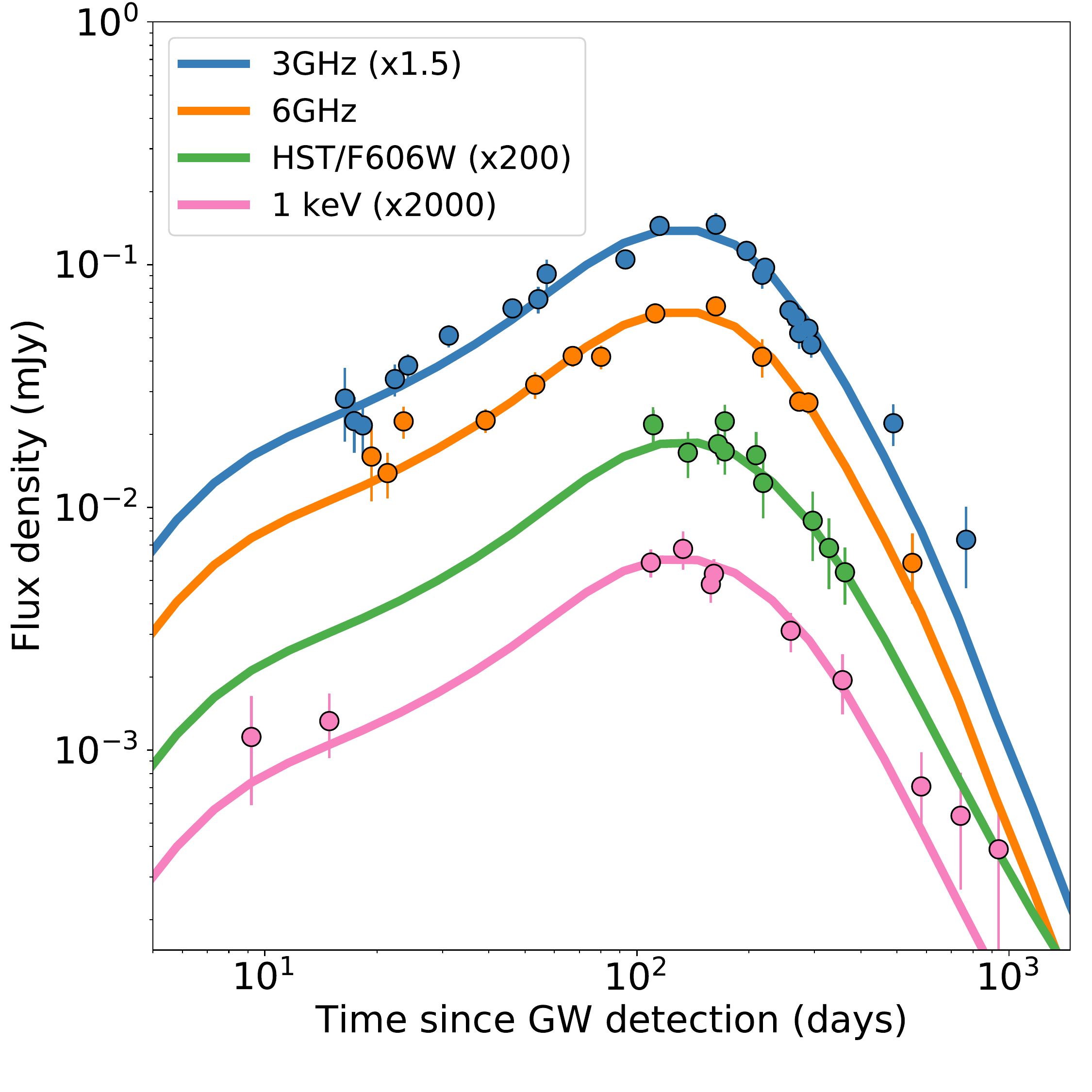}\\
\includegraphics[width=3.5in]{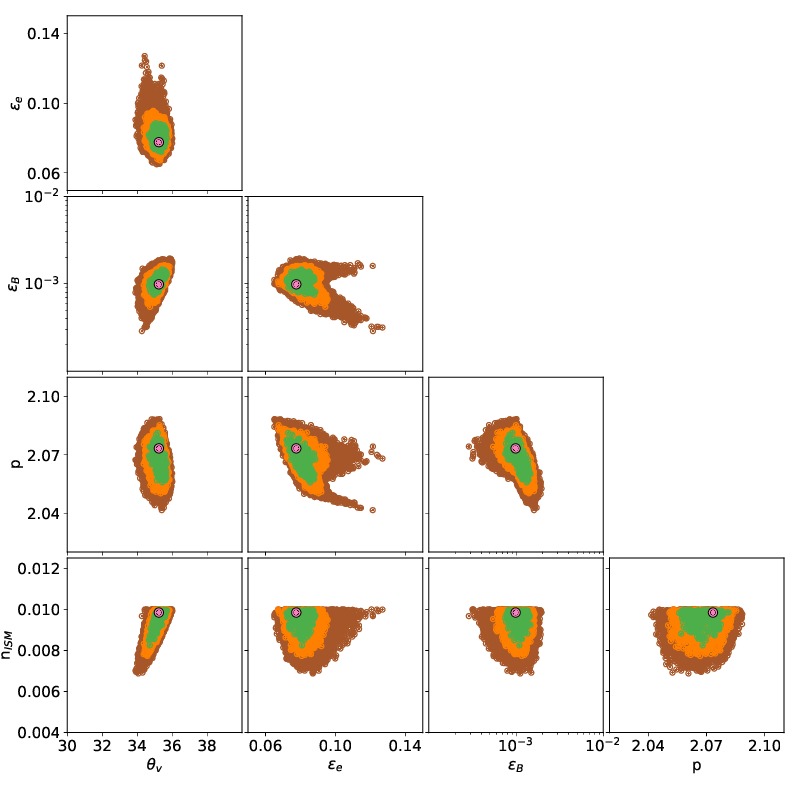}\\
\caption{Upper panel: Structured jet model fit to the afterglow light curve using the forward-shock model from \cite{lazzati2018}. The best-fit model from the MCMC analysis is shown and color coding is according to the observing frequency.
Only a subset of the observational data is plotted in this figure.
Lower panel: Corner plot for the structured jet model.
The pink dot shows the best-fit value for each parameter, and green indicates the 67\% confidence interval.
See \S\ref{sec:literature:sj} for details.}
\label{fig:structure_jet_lc}
\end{figure}

We modeled the data with a forward shock afterglow model based on the semi-analytic code used in \cite{lazzati2018}.
The model has 5 free parameters: the viewing angle $\theta_{\rm v}$, the microphysical parameters $\epsilon_e$ (the fraction of shock energy given to electrons) and $\epsilon_B$ (the fraction of shock energy given to tangled magnetic field), the electron’s population distribution index $p$, and the external medium density $n_{\rm ISM}$, which was assumed to be constant. The total kinetic energy of the fireball and its initial Lorentz factor, both dependent on the viewing angle, were taken from an hydrodynamic numerical simulation, previously described in \cite{lazzati2017}. Specifically, the energy of the blast wave, set by the numerical simulation, is $6\times 10^{49}$\,erg. The fit was performed with a dedicated implementation of a Markov-Chain Monte Carlo scheme, assuming flat priors for all free parameters except the interstellar medium density, which was assumed to be small \citep[$n_{\rm ISM}\le0.01\text{~cm}^{-3}$; see \S\ref{sec:obs-constraints} and][]{hallinan2017}. 
The best fit parameters are give in Table~\ref{tab:best-fit}.
The fit to the light curve and corresponding corner plot are shown in Figure~\ref{fig:structure_jet_lc}.

We caution that the quoted uncertainties in the best fit results are purely statistical and do not reflect the potentially large systematic uncertainties associated with the numerical simulation itself. We have modeled the jet dynamics and merger ejecta interaction as a purely hydrodynamical system using FLASH \citep{2000ApJS..131..273F}, which does not have magnetohydrodynamic capabilities for relativistic flows. This can affect the final polar distribution of both the kinetic energy of the merger and the Lorentz factor of the outflow. Additionally, the FLASH simulation uses a set of initial conditions that do not necessarily reproduce the actual conditions at the base of the jet \citep{lazzati2017}.  Finally, it should be noted that the constraint on the proper motion of the radio transient \citep{mooley2018-vlbi} are not included in the fit\footnote{Our code is currently being updated to take the proper motion constraint into account.}.

\subsection{Review of and comparison with previous models}\label{sec:literature:review}

It is now widely accepted that \object{GW170817} produced a relativistic jet with substantial angular structure. 
Multiple studies, using different hydrodynamic and semi-analytic models, have modeled the afterglow of \object{GW170817} to obtain the properties of the structured jet. 
The parameter constraints derived by these studies are tabulated in Table~\ref{tab:modeling_summary}. 
Generally, all studies obtain small jet opening angles, $\lesssim6\degr$, and viewing angles between 15--35$\degr$.
Very few studies \citep{mooley2018-vlbi,hotokezaka2019,ghirlanda2019} fit the VLBI proper motion constraint.
Estimates for the isotropic equivalent energy and circummerger density range from  $10^{51.5}$--$10^{53}$\,erg and 
 $10^{-1.5}$--$10^{-4.5}$\,cm$^{-3}$, respectively.  The circummerger density is correlated with $\epsilon_B$, which is estimated to be between 10$^{-1.5}$--$10^{-5}$.
The parameters we obtain using the structure jet model are generally in agreement with the literature, although our estimate of $\theta_v = 35.2\degr\pm0.6\degr$ is the largest to-date.

\cite{resmi2018,hotokezaka2019,troja2019,ghirlanda2019,lamb2019,ryan2020} use semi-analytical techniques, using power-law and/or Gaussian angular profiles for the jet.
\cite{lamb2019} also use a two-component model that includes a top-hat jet and Gaussian cocoon. 
Only \cite{hotokezaka2019,ghirlanda2019} include VLBI constraints (but these studies also neglect sideways/lateral expansion of the jet).
This may be the reason why they obtain low viewing angles, $\sim 16\degr$.
The other semi-analytical studies obtain median viewing angle estimates between $20\degr$--$27\degr$.

\cite{wu2019,hajela2019} fit a ``boosted  fireball" model, a family of models parameterized by two parameters viz. internal energy and boost Lorentz factor (which also define the angular structure of the structured jet), to the afterglow light curve. 
This technique uses a template bank constructed from 3D hydrodynamical simulations together with analytic scaling relations. 
\cite{lazzati2018} use input data (jet energy and opening  angle, chosen to mimic a typical SGRB) from a 3D/2D hydrodynamical simulation.
Sideways expansion is neglected.
This work also utilizes the \cite{lazzati2018} scheme.
All of these studies obtain median viewing angles larger than $30\degr$.
\cite{mooley2018-vlbi} carry out a dozen hydrodynamical simulations using various setups to determine the parameters that can fit the afterglow light curve and proper motion data.
They obtain a median viewing angle of about $20\degr$, which is much smaller than that obtained by \cite{lazzati2018,wu2019,hajela2019}, likely due to the VLBI proper motion being taken into account.

Looking at the substantial differences in all these modeling results, we therefore highlight that a combined analysis of the VLBI proper motion measured by \cite{mooley2018-vlbi} and of the full uniform light curve presented here is crucial for obtaining an accurate estimate of the viewing angle and other parameters (jet and ISM) of \object{GW170817}.

\section{Summary \& Discussion}\label{sec:discussion}

By compiling and reprocessing archival data at radio, optical and X-ray wavelengths, and reporting new radio data, we have presented a fairly uniform dataset of the afterglow of \object{GW170817} between 0.5--940 days post-merger.
These afterglow measurements are available in ASCII format (continuously updated as new data get published) on the web\footnote{\url{https://github.com/kmooley/GW170817/} or \url{http://www.tauceti.caltech.edu/kunal/GW170817/}}.
The afterglow light curve (Figure~\ref{fig:lc}) shows a power-law rise, $F_\nu \propto t^{0.86}$ and a power-law decline $F_\nu \propto t^{-1.92}$, consistent with expectations for a laterally expanding relativistic jet core (dominating the late-time afterglow emission) surrounded by low-Lorentz factor material (dominating the early-time afterglow emission).
A more detailed investigation of the shape of the afterglow light curve may provide important insights into the angular structure of the relativistic outflow from \object{GW170817}, and possibly the properties of the ejecta and jet at the time of jet launch.

Our uniform panchromatic dataset of the afterglow implies a spectral index of $\beta=-0.584\pm0.002$, leading to an extremely precise estimate of the electron power-law index, $p=2.168\pm0.004$.
The single unchanging spectral index across all epochs implies that the synchrotron self-absorption frequency is below the radio band and the cooling frequency is above the soft X-ray band throughout the evolution of the afterglow.
The rate of decline of the afterglow light curve appears to be consistent with $t^{-p}$ (within 2$\sigma$), indicating that sound-speed expansion may provide a sufficiently accurate description of the jet lateral expansion.
We do not find any evidence for steepening beyond $t^{-p}$ post-peak, as found by some hydrodynamical simulations \citep[e.g.][]{vanEerten2013}.

Likewise, we do not find any evidence for flattening of the light curve during the decline phase.
Specifically, we do not see any component declining as $t^{-1}$, even during the most recent radio observing epoch (767 days post-merger), that may be expected for a cocoon.
Interpolating a $t^{-1}$ power law backwards in time from our 3 GHz detection 767 days post-merger, we find that any cocoon contribution to the afterglow is negligible beyond $\sim$90 days post-merger, i.e. the jet core has likely dominated the afterglow beyond 90 days post-merger. 
Even at 900 days post-merger, the lack of any flattening in the light curve indicates that the jet is still in the relativistic phase, $\Gamma\beta\gtrsim1$.
This also implies that the counter-jet and the late-time dynamical ejecta afterglow have not made their appearance yet.

In our reprocessing of the X-ray data, we do not find any evidence for significant flaring or synchrotron cooling, which argues against the presence of any long-lived magnetized neutron star remnant in \object{GW170817} \citep[see also][]{lyman2018,piro2019,hajela2019}.
By analyzing the Chandra data we also study the diffuse X-ray emission of \object{NGC 4993}. 
The X-ray luminosity in $0.5$--$8$\,keV is estimated about $2\times 10^{39}$\,erg\,s$^{-1}$, somewhat lower than the X-ray luminosity of early-type galaxies \citep{Babyk2018ApJ} given the total mass of \object{NGC 4993} of $\sim 10^{12}M_{\odot}$ \citep{Pan2017ApJ}.
We estimate the hot ionized ISM density from the observed X-ray flux to be $\lesssim 0.01\,{\rm cm^{-3}}$. This result is consistent with that obtained from an independent analysis done by \cite{hajela2019} as well as that estimated from the afterglow modelings of \object{GW170817}.
Comparing the late-time X-ray luminosity with the pulsar spin-down luminosity, we rule out the phase space of $10^{10.4}\sim10^{16}$~G for the magnetic field strength of any possible long-lived neutron star remnant in GW170817. 

We fit both analytic and hydrodynamical models to the non-thermal afterglow to estimate the physical, geometrical and microphysical parameters associated with the jet.
These parameters are tabulated in Table~\ref{tab:best-fit}.
We find that the ratio of the jet energy to the ISM density, $E/n_{\rm ISM}$, to be  $\mathcal{O}(52)$--$\mathcal{O}(53)$.
The jet opening angle is a few degrees for all the models. The best-fit viewing angle is calculated to be $\simeq15\degr$  using analytical models that include the VLBI proper motion constraint, but $\simeq35\degr$ when using a structured jet model that does not attempt to fit the VLBI proper motion constraint. 
These viewing angle estimates lie at the tail end of the values previously published.

The wide range of viewing angles $\theta_v$, seemingly inconsistent within uncertainties (see Table~\ref{tab:modeling_summary}) obtained using different modeling techniques (used here and in previous studies), can be explained through the fitting of the VLBI proper motion measurement \citep{mooley2018-vlbi} and the systematic uncertainties (which may not be quantifiable) associated with the models/simulations.
Fitting of the VLBI proper motion is especially important since the viewing angle cannot be estimated accurately by the light curve alone. 
As long as $\theta_j<\theta_v\ll1$, the light curve depends only on the ratio $\theta_v/\theta_j$ and not on $\theta_v$ or $\theta_j$ alone. 
Once the proper motion measurement is taken into account, it breaks the degeneracy and $\theta_v$ can be constrained \citep{mooley2018-vlbi,mooley2018-strongjet,nakar2020}.
We highlight the importance of considering these limitations when drawing any physical conclusions or comparing the modeling results obtained using different techniques.


Given the upper limit on the ISM density, $\lesssim 0.01\,{\rm cm^{-3}}$ (\S\ref{sec:obs-constraints}), we predict that the radio emission arising from the dynamical ejecta (kilonova ejecta) can be detected in the future \citep{nakarpiran2011, alexander2017,radice2018,hajela2019,kathirgamaraju2019},
depending on the actual ISM density and the velocity distribution of the ejecta. The latter is quite sensitive to the neutron equation of state (EOS; e.g., \citealt{hotokezaka2018ApJ,radice2018}).
For example, a soft EOS predicts radio remnant with a flux density at $\sim 10\,\mu{\rm Jy}$ level on a time scale of $10$\,yr, if the ISM density is $\sim 10^{-3}\,{\rm cm^{-3}}$ \citep{radice2018}. Detecting a long-lasting radio remnant will offer an opportunity to constrain the neutron star EOS from the light curve.







\quad \newline
Acknowledgements:
The authors thank Eleonora Troja and Brendan Connor for help with correcting the X-ray flux densities, Schuyler van Dyk for helpful discussions on HST data, and the anonymous referee for detailed comments that helped improve the manuscript.
The MeerKAT telescope is operated by the South African Radio Astronomy Observatory, which is a facility of the National Research Foundation, an agency of the Department of Science and Innovation.
The National Radio Astronomy Observatory is a facility of the National Science Foundation operated under cooperative agreement by Associated Universities, Inc. 
The authors thank the NRAO staff, especially Mark Claussen and Amy Mioduszewski, for scheduling the VLA observations. 
The Australia Telescope is funded by the Commonwealth of Australia for operation as a National Facility managed by CSIRO. We acknowledge the Gomeroi people as the traditional owners of the Observatory site.
We thank the staff of the GMRT that made these observations possible. GMRT is run by the National Centre for Radio Astrophysics of the Tata Institute of Fundamental Research.
KPM and GH acknowledge support from the National Science Foundation Grant AST-1911199.
DD is supported by an Australian Government Research Training Program Scholarship. TM acknowledges the support of the Australian Research Council through grant DP190100561.
Parts of this research were conducted by the Australian Research Council Centre of Excellence for Gravitational Wave Discovery (OzGrav), project number CE170100004. We acknowledge support by the GROWTH (Global Relay of Observatories Watching Transients Happen) project funded by the National Science Foundation PIRE (Partnership in International Research and Education) program under Grant No 1545949.
DL acknowledges support from NASA grants 80NSSC18K1729 (Fermi) and NNX17AK42G (ATP), Chandra grant TM9-20002X, and NSF grant AST-1907955.
This research has made use of NASA's Astrophysics Data System Bibliographic Services.
CF gratefully acknowledges support of his research by the Heising-Simons Foundation.
JM acknowledges financial support from the State Agency for Research of the Spanish MCIU through the ``Center of Excellence Severo Ochoa'' award to the Instituto de Astrof\'isica de Andaluc\'ia (SEV-2017-0709) and from the grant  RTI2018-096228-B-C31 (MICIU/FEDER, EU). DLK was supported by NSF grant AST-1816492
PK is partially supported by the BMBF project 05A17PC2for D-MeerKAT.
A.B. and A.C. acknowledge support from the National Science Foundation via grant \#1907975.

\facility{VLA, ATCA, uGMRT, MeerKAT, eMERLIN, HST, Chandra, XMM-Newton}
\software{
{\tt MIRIAD}~\citep{sault1995},~{\tt CASA} \citep[release 5.6.2-3.el7]{mcmullin2007},~{\tt  emcee}~\citep{foreman-mackey2013},~{\tt corner}~\citep{foreman-mackey2016}}, {\tt CARACal}~\citep[beta version ]{ramatsoku2020}, {\tt AOFlagger} ~\citep[version 2.14.0]{offriga2010}, {\tt WSClean}~\citep[version 2.8.0]{offringa2014}, {\tt CubiCal}~\citep[version 1.2.2]{kenyon2018}, {\tt MeqTrees}~\citep[version 1.6.0]{noordam2010}

\clearpage
\begin{center}
\begin{longtable*}{lllllllp{5cm}}
\caption{Radio Afterglow Measurements of GW170817. } \label{tab:alldata} \\

\hline\hline \multicolumn{1}{c}{UT date} & \multicolumn{1}{c}{$\Delta$T$^{\dag}$} & \multicolumn{1}{l}{Telescope} & \multicolumn{1}{c}{$\nu$} & \multicolumn{1}{c}{$F_\nu$} & \multicolumn{1}{c}{$\sigma_\nu$} & \multicolumn{1}{c}{Reproc?} & \multicolumn{1}{l}{Original Reference} \\ 
\multicolumn{1}{c}{ } & \multicolumn{1}{l}{(d)} & \multicolumn{1}{l}{} & \multicolumn{1}{l}{(GHz)} & \multicolumn{1}{l}{($\mu$Jy)} & \multicolumn{1}{l}{($\mu$Jy)} & \multicolumn{1}{l}{} \\ \hline  
\endfirsthead

\multicolumn{7}{c}
{{\bfseries \tablename\ \thetable{} -- continued from previous page}} \\
\hline \multicolumn{1}{c}{UT date} & \multicolumn{1}{l}{$\Delta$T$^{\dag}$} & \multicolumn{1}{l}{Telescope} & \multicolumn{1}{l}{$\nu$} & \multicolumn{1}{l}{$F_\nu$} & \multicolumn{1}{l}{$\sigma_\nu$}  & \multicolumn{1}{l}{Reproc?} & \multicolumn{1}{l}{Original Reference} \\ 
\multicolumn{1}{l}{ } & \multicolumn{1}{l}{(d)} & \multicolumn{1}{l}{} & \multicolumn{1}{l}{(GHz)} & \multicolumn{1}{l}{($\mu$Jy)} & \multicolumn{1}{l}{($\mu$Jy)} & \multicolumn{1}{l}{} \\ \hline  
\endhead
\hline \multicolumn{7}{r}{{Continued on next page}} \\ \hline
\endfoot
\hline \hline
\endlastfoot
    \hline 
    2017 Aug 18.10 & 0.57   & VLA   &  9.7   & $<144$   & \nodata  & N  &    \cite{alexander2017} \\
    2017 Aug 18.21 & 0.68 & ATCA & 8.5 & $<120$ & \nodata & N  &  \cite{hallinan2017}  \\
    2017 Aug 18.21 & 0.68 & ATCA & 10.5 & $<150$ & \nodata & N  &  \cite{hallinan2017}  \\
    2017 Aug 18.46 & 0.93 & uGMRT & 0.61 & $<195$ & \nodata & N  &  \cite{hallinan2017}  \\
    2017 Aug 18.95 & 1.43 & ALMA & 338.5 & $<126$ & \nodata & N  &  \cite{kim2017} \\
    2017 Aug 18.97 & 1.44   & VLA   &  10.0  & $<13.8$  & \nodata  & N  &    \cite{alexander2017} \\
    2017 Aug 18.97 & 1.44 & VLITE\footnote{VLITE is the VLA Low
    Band Ionosphere and Transient Experiment}/VLA & 0.34 & $<34800$ & \nodata & N  &  \cite{hallinan2017} \\
    2017 Aug 19.95 & 2.41   & ALMA  &  97.5  & $<75$   & \nodata  & N  &    \cite{alexander2017} \\
    2017 Aug 19.95 & 2.42   & VLA   &  15.0  & $<17.7$  & \nodata  & N  &    \cite{alexander2017} \\
    2017 Aug 19.95 & 2.43   & VLA & 6.2 & $<20$ & \nodata  & N  & \cite{hallinan2017} \\
    2017 Aug 19.95 & 2.43   & VLA & 9.7 & $<17$ & \nodata  & N  & \cite{hallinan2017} \\
    2017 Aug 19.95 & 2.43   & VLA & 15 & $<22$ &  \nodata & N  & \cite{hallinan2017} \\
    2017 Sep 19.97 & 2.44   & VLITE/VLA & 0.34 & $<28000$ & \nodata & N  &  \cite{hallinan2017} \\
    2017 Aug 19.97 & 2.44   & VLA   &  10.0  & $<17.1$  & \nodata  & N  &    \cite{alexander2017} \\
    2017 Aug 19.97 & 2.46   & VLA   &  6.0   & $<21.9$  & \nodata  & N  &    \cite{alexander2017} \\
    2017 Aug 20.31 & 2.78   & uGMRT & 0.4 & $<780$ & \nodata &  N &  \cite{hallinan2017} \\
    2017 Aug 20.46 & 2.93   & uGMRT & 1.2 & $<98$ & \nodata & N  &  \cite{hallinan2017} \\
    2017 Aug 20.76 & 3.23   & ALMA & 338.5 & $<90$ & \nodata & N  &  \cite{kim2017} \\
    2017 Aug 20.87 & 3.34   & VLA & 3 & $<32$ & \nodata & N  & \cite{hallinan2017} \\
    2017 Aug 20.87 & 3.34   & VLITE/VLA & 0.34 & $<44700$ & \nodata & N  &  \cite{hallinan2017} \\
    2017 Aug 21.23 & 3.67   & ATCA & 8.5 & $<135$ & \nodata & N  & \cite{hallinan2017} \\
    2017 Aug 21.23 & 3.67   & ATCA & 10.5 & $<99$ & \nodata & N  &  \cite{hallinan2017} \\
    2017 Aug 23.0 & 5.48   & VLA   &  10.0  & $<28.5$  & \nodata  & N  &    \cite{alexander2017} \\
    2017 Aug 25.4 & 7.9     & uGMRT & 1.39  & $<69$   &  \nodata  & N    &  \cite{kim2017} \\
    2017 Aug 25.8 & 8.29   & VLA   &  10.0  & $<17.4$  & \nodata  & N  &    \cite{alexander2017} \\
    2017 Aug 25.96 & 8.37 & VLITE/VLA & 0.34 & $<37500$ & \nodata & N  &  \cite{hallinan2017} \\
    2017 Aug 25.96 & 8.43 & ALMA & 338.5 & $<150$ & \nodata & N  &  \cite{kim2017} \\
    2017 Aug 26.96 & 9.43 & ALMA & 338.5 & $<102$ & \nodata &  N &  \cite{kim2017} \\
    2017 Aug 27.00 & 9.43   & ALMA  &  97.5  & $<72$   & \nodata  & N  &    \cite{alexander2017} \\
    2017 Aug 28.2 & 10.6 & ATCA & 8.5 & $<54$ & \nodata &  N &  \cite{hallinan2017} \\
    2017 Aug 28.2 & 10.6 & ATCA & 10.5 & $<39$ & \nodata & N  &  \cite{hallinan2017} \\
    2017 Aug 29.5 & 11.9 & uGMRT & 0.7 & $<123$ & \nodata & N  &  \cite{hallinan2017} \\
    2017 Aug 30.9 & 13.4  & VLA   &  10.0  & $<18.3$  & \nodata  & N  &    \cite{alexander2017} \\
    2017 Aug 31.0 & 13.5 & VLITE/VLA & 0.34 & $<20400$ & \nodata & N  &  \cite{hallinan2017} \\
    2017 Aug 31.0 & 13.5 & VLA & 6.2 & $<17$ & \nodata & Y  &  \cite{hallinan2017} \\
    2017 Aug 31.5 & 13.9 & uGMRT & 0.4 & $<600$ & \nodata & N  &  \cite{hallinan2017} \\
    2017 Sep 1.8  & 15.3  & ALMA  &  97.5  & $<39$   &  \nodata &  N &    \cite{alexander2017} \\
    2017 Sep 1.9 & 15.4 & VLA & 6.2 & $<13$ & \nodata &  N &  \cite{hallinan2017} \\
    2017 Sep 1.9 & 15.4 & VLITE/VLA & 0.34 & $<11400$ & \nodata & N  &  \cite{hallinan2017} \\
    2017 Sep 2.9 & 16.4 & VLITE/VLA & 0.34 & $<11700$ & \nodata & N  &  \cite{hallinan2017} \\
    2017 Sep 2.9 & 16.4 & VLA & 3 & 18.7 & 6.3 & N  &  \cite{hallinan2017} \\
    2017 Sep 3.0 & 16.5 & VLA & 6.2 & $<15$ & \nodata &  Y &  \cite{hallinan2017} \\
    2017 Sep 3.9 & 17.4 & VLA & 3 & 15.1 & 3.9 & N  &  \cite{hallinan2017} \\
    2017 Sep 3.9 & 17.4 & VLITE/VLA & 0.34 & $<6900$ & \nodata & N  &  \cite{hallinan2017} \\
    2017 Sep 4.0 & 17.5 & VLA & 6.2 & $<15$ & & \nodata & This work \\
    2017 Sep 4.9 & 18.3 & VLA & 3 & 14.5 & 3.7 & N  &  \cite{hallinan2017} \\
    2017 Sep 5.2 & 18.7 & ATCA & 7.25 & 15.4 & 4.8 & Y  &  \cite{hallinan2017},\cite{mooley2018-strongjet} \\
    2017 Sep 5.5 & 19.0 & uGMRT & 0.7 & $<140$ & \nodata & N  &  \cite{hallinan2017} \\
    2017 Sep 5.9 & 19.4 & VLA & 6.2 & 15.9 & 5.5 & N &  \cite{hallinan2017} \\
    2017 Sep 5.9  & 19.4  & VLA   &  10.0  & $<13.5$  &  \nodata & N  &    \cite{alexander2017} \\
    2017 Sep 5.9  & 19.4  & VLA   &  6.0   & 19  & 6  & N  &    \cite{alexander2017} \\ 
    2017 Sep 6.0 & 19.5 & VLA & 10 & $<14$ & \nodata & \nodata & This work \\
    2017 Sep 7.9 & 21.4 & VLITE/VLA & 0.34 & $<8100$ & \nodata  & N  &  \cite{hallinan2017} \\\
    2017 Sep 7.9 & 21.4 & VLA & 6.2 & 13.6 & 2.9 & N  &  \cite{hallinan2017} \\
    2017 Sep 8.9 & 22.4 & VLA & 3 & 22.5 & 3.4 & N  &  \cite{hallinan2017} \\
    2017 Sep 8.9 & 22.4 & VLITE/VLA & 0.34 & $<6300$ & \nodata &  N &  \cite{hallinan2017} \\
    2017 Sep 9.4 & 23.0    & uGMRT & 1.39  & $<108$   & \nodata   &  N  &  \cite{kim2017} \\
    2017 Sep 9.9 & 23.4 & VLITE/VLA & 0.34 & $<4800$ & \nodata & N  &  \cite{hallinan2017} \\
    2017 Sep 9.9 & 23.4 & VLA & 6 & 22.6 & 3.4 & Y & \cite{hallinan2017} \\
    2017 Sep 10.8 & 24.2 & VLA & 3 & 25.6 & 2.9 & N  &  \cite{hallinan2017} \\
    2017 Sep 10.9 & 24.3 & VLITE/VLA & 0.34 & $<6600$ & \nodata & N  &  \cite{hallinan2017} \\
    2017 Sep 16.3 & 29.8 & uGMRT & 1.39  & $<126$   & \nodata &  N  &  \cite{kim2017} \\
    2017 Sep 16.3 & 29.7 & uGMRT & 0.68 & $<246$ & \nodata & N  &  \cite{mooley2018-wideoutflow} \\
    2017 Sep 16.9 & 30.3  & ALMA  &  97.5  & $<42$   & \nodata  & N  &    \cite{alexander2017} \\
    2017 Sep 17.8 & 31.3 & VLA & 3 & 34 & 3.6 & N  &  \cite{mooley2018-wideoutflow} \\
    2017 Sep 21.9 & 35.3 & VLA & 1.5 & 44 & 10 & N  &  \cite{mooley2018-wideoutflow} \\
    2017 Sep 23-24 & 36.9  & VLA & 1.6 & $<40$ & \nodata & N & \cite{mooley2018-vlbi} \\
    2017 Sep 25.8 & 39.2 & VLA & 6 & 22.8 & 2.6 & Y  &  \cite{alexander2017} \\
    2017 Sep 26.0 & 39.4 & VLA & 15 & $<18$ & \nodata & \nodata  &  This work \\
    2017 Sep 30.0 & 44.1 & ALMA & 338.5 & $<93$ & \nodata & N & \cite{kim2017} \\
    2017 Oct 2.8 & 46.3 & VLA & 3 & 44 & 4 & N  & \cite{mooley2018-wideoutflow} \\
    2017 Oct 7-8 & 51.5 & VLA & 3.2 &$<60$  & \nodata & N  & \cite{mooley2018-vlbi} \\
    2017 Oct 9.8 & 53.3 & VLA & 6 & 32 & 4 & N  &  \cite{mooley2018-wideoutflow} \\
    2017 Oct 10.8 & 54.3 & VLA & 3 & 48 & 6 & N  &  \cite{mooley2018-wideoutflow} \\
    2017 Oct 13.7 & 57.2 & VLA & 3 & 61 & 9 & N  &  \cite{mooley2018-wideoutflow} \\
    2017 Oct 20--26 & 65.9 & uGMRT & 0.67 & 148 & 22 & Y  & \cite{mooley2018-wideoutflow}\\
    2017 Oct 23.3 & 66.6 & uGMRT & 1.3 & 98 & 20 & Y & \cite{resmi2018} \\
    2017 Oct 23.7 & 67.2 & VLA & 6 & 42.6 & 4.1 & N  &  \cite{mooley2018-wideoutflow} \\
    2017 Oct 28-Nov 4 & 72.2 & VLA & 4.5 & 58 & 5 & N  &  \cite{mooley2018-vlbi} \\
    2017 Nov 1.0 & 75.5 & ATCA & 7.35 & 35.9 & 4.3 & N  & \cite{mooley2018-wideoutflow},\cite{mooley2018-strongjet} \\
    2017 Nov 3.1 & 77.6 & uGMRT & 1.4 & 97 & 16 & Y & \cite{resmi2018} \\
    2017 Nov 4.7 & 79.2 & VLA & 4.5 & 45 & 7 & N  &  \cite{mooley2018-vlbi} \\
    2017 Nov 5.7 & 80.1 & VLA & 6 & 41.7 & 4.7 & \nodata & This work  \\
    2017 Nov 17.9 & 92.4 & ATCA & 7.25 & 31.7 & 4.3 & N  &  \cite{ddobie2018}, \cite{mooley2018-strongjet} \\ 
    2017 Nov 17.9 & 75.5 & ATCA & 7.35 & 39.6 & 7 & N  & \cite{mooley2018-wideoutflow} \\
    2017 Nov 18.6 & 93.1 & VLA & 1.5 & 98 & 14 & N   &  \cite{mooley2018-wideoutflow} \\
    2017 Nov 18.6 & 93.1 & VLA & 3 & 70 & 5.7 & N  &  \cite{mooley2018-wideoutflow} \\
    2017 Nov 18.7 & 93.2 & VLA & 15 & 26 & 4.4 &  Y &  \cite{mooley2018-wideoutflow}\\
    2017 Nov 20--27 & 97.1 & uGMRT & 0.67 & 199 & 16 & Y  & \cite{resmi2018}\\
    2017 Dec 2 & 107 & uGMRT & 1.3 & 141 & 20 & Y &  \cite{resmi2018} \\
    2017 Dec 2 & 107 & ATCA & 1.3 & 53.2 & 4.5 & N &  \cite{mooley2018-strongjet} \\
    2017 Dec 7 & 112 & VLA & 6 & 62.9 & 3.2 & Y & \cite{margutti2018} \\
    2017 Dec 10 & 115 & VLA & 3 & 96.2 & 8 & Y & \cite{margutti2018} \\
    2017 Dec 10 & 115 & VLA & 10 & 51.2 & 3.4 & Y & \cite{margutti2018} \\
    2017 Dec 10 & 115 & VLA & 15 & 41.2 & 1.9 & Y & \cite{margutti2018} \\
    2017 Dec 20 & 125 & ATCA & 7.25 & 58.2 & 5 & N  &  \cite{ddobie2018}, \cite{mooley2018-strongjet} \\
    2017 Dec 20 & 125 & uGMRT & 1.3 & 149 & 17 & Y & \cite{resmi2018} \\
    2017 Dec 21 & 126 & uGMRT & 0.67 & 221 & 19 & Y & \cite{resmi2018} \\
    2017 Dec 25-Jan 2 & 134 & LOFAR & 0.114 & $<6300$ & \nodata & N & \cite{broderick2020} \\
    2018 Jan 13 & 149 & ATCA & 7.25 & 60.6 & 4.3 & N  &  \cite{ddobie2018}, \cite{mooley2018-strongjet} \\
    2018 Jan 14 & 150 & eMERLIN & 5.1 & 90 & 30 & \nodata  &  This work \\
    2018 Jan 16 & 152 & uGMRT & 1.3 & 171 & 18 & Y & \cite{resmi2018}\\
    2018 Jan 20 & 155 & MeerKAT & 1.3 & 151 & 23 & 
\nodata & This work \\
    2018 Jan 27 & 163 & VLA & 3 & 97.3 & 11.3 & Y & \cite{margutti2018} \\
    2018 Jan 27 & 163 & VLA & 6 & 67.3 & 4.1 & Y & \cite{margutti2018} \\
    2018 Jan 27 & 163 & VLA & 10 & 47.4 & 3.6 & Y & \cite{margutti2018} \\
    2018 Jan 27 & 163 & VLA & 15 & 39.6 & 2 & Y & \cite{margutti2018} \\
    2018 Feb 1 & 167 & ATCA & 7.25 & 57.9 & 6.9 & N  &  \cite{ddobie2018}, \cite{mooley2018-strongjet} \\
    2018 Feb 17 & 183 & uGMRT & 0.65 & 211 & 34 & N  &  \cite{mooley2018-strongjet} \\
    2018 Feb 13-28 & 187 & eMERLIN & 5.1 & $<83$ & \nodata & \nodata  &  This work \\
    2018 Mar 2 & 197 & VLA & 3 & 75.9 & 5.2 & N  &  \cite{ddobie2018} \\
    2018 Mar 3 & 197 & MeerKAT & 1.3 & 107 & 17 & Y & \cite{mooley2018-strongjet} \\
    2018 Mar 01--06 & 198 & eMERLIN & 5.1 & $<90$ & \nodata & \nodata  &  This work \\
    2018 Mar 12-13 & 207 & gVLBA & 5 & 42 & 12 & N & \cite{ghirlanda2019}\\
    2018 Mar 8--22 & 210 & eMERLIN & 5.2 & $<$60 & \nodata & N  &  \cite{ghirlanda2019} \\
    2018 Mar 21 & 216 & VLA & 10 & 36.3 & 3.6 & N  &  \cite{mooley2018-strongjet} \\
    2018 Mar 22 & 217 & VLA & 3 & 60.5 & 7.5 & Y & \cite{alexander2018} \\
    2018 Mar 22 & 217 & VLA & 6 & 41.7 & 7.5 & Y & \cite{alexander2018} \\
    2018 Mar 22 & 217 & VLA & 10 & 32.6 & 4 & Y & \cite{alexander2018} \\
    2018 Mar 22 & 217 & VLA & 15 & 24.7 & 3.1 & Y & \cite{alexander2018} \\
    2018 Mar 25-26 & 218 & VLA & 3 & 64.7 & 2.7 & N  &  \cite{mooley2018-strongjet} \\ 
    2018 Mar 27 & 222 & ATCA & 7.25 & 39.7 & 7.2 & N  &  \cite{mooley2018-strongjet} \\ 
    2018 Apr 1-10 & 229 & VLA & 4.5 & 48 & 6 & N  &  \cite{mooley2018-vlbi} \\
    2018 Apr 26 & 252 & MeerKAT & 1.3 & 74 & 9 & \nodata  &   This work \\ 
    2018 May 1 & 257 & VLA & 3 & 43.2 & 5.8 & Y & \cite{alexander2018} \\
    2018 May 6 & 261 & MeerKAT & 1.3 & 66 & 10 & \nodata  &   This work \\
    2018 May 11-12 & 267 & ATCA & 7.25 & 25 & 4.1 & N  &  \cite{mooley2018-strongjet} \\
    2018 May 12 & 267 & VLA & 3 & 40.3 & 2.7 & N  &  \cite{mooley2018-strongjet} \\
    2018 May 17 & 273 & VLA & 3 & 34.8 & 4.9 & Y & \cite{alexander2018} \\ 
    2018 May 17 & 273 & VLA & 6 & 27.2 & 2.1 & Y & \cite{alexander2018} \\ 
    2018 May 13-25 & 275 & uGMRT & 0.65 & $<153$ & \nodata & N & \cite{mooley2018-strongjet} \\
    2018 Jun 2 & 289 & VLA & 3 & 36.3 & 3.9 & Y & \cite{alexander2018} \\ 
    2018 Jun 2 & 289 & VLA & 6 & 27 & 2.8 & Y & \cite{alexander2018} \\ 
    2018 Jun 7 & 294 & VLA & 3 & 31.2 & 3.6 & N  &  \cite{mooley2018-strongjet} \\
    2018 Jun 11 & 298 & ATCA & 7.25 & 23.4 & 4.2 & N  &  \cite{mooley2018-strongjet} \\
    2018 Jul 4 & 320 & ATCA & 7.25 & 23.1 & 4.0 & Y & \cite{troja2019}\\
    2018 Jul 8 & 324 & MeerKAT & 1.3 & 47.2 & 12.8 & \nodata  &  This work \\
    2018 Aug 12 & 359 & ATCA & 7.25 & 15.5 & 5.0 & Y & \cite{troja2019}\\
2018 Aug 23-26 & 372 & LOFAR & 0.114 & $<18600$ & \nodata & N & \cite{broderick2020} \\
    2018 Sep 2 & 380 & MeerKAT & 1.3 & 37.9 & 11.1 & \nodata  &  This work \\
    2018 Sep 13 & 391 & ATCA & 7.25 & $<13$ & \nodata & Y & \cite{troja2019}\\
    2018 Nov 21 & 461 & ATCA & 7.25 & $<11$ & \nodata & \nodata & This work\\
    2018 Dec 18--20	 & 489 & VLA & 3 & 14.8 & 2.9 & \nodata & This work \\ 
    2019 Dec 20-Jan 3 &  496 & MeerKAT & 1.3  & $<22$ & \nodata & \nodata & This work \\ 
    2019 Jan 14 & 515 & ATCA & 7.25 & $<13$ & \nodata & \nodata & This work\\
    2019 Jan 21--Mar 29 & 545 & VLA & 6 & 5.9 & 1.9 & N & \cite{hajela2019}\\
    2019 Mar 19 & 580 & ATCA & 7.25 & $<18$ & \nodata & \nodata & This work (PI: Piro)\\
    2019 Aug 11--30 & 734 & VLA & 6 & $<$8.4 & \nodata & N & \cite{hajela2019} \\ 
    
    2019 Sep 16 & 760 & ATCA & 7.25 & $<13$ & \nodata & \nodata & This work\\
    2019 Sep 21--27	 & 767 & VLA & 3 & 4.9 & 1.8 & \nodata & This work \\
    \hline
    \hline
    \multicolumn{8}{c}{X-ray measurements} \\ \hline
    2017 Aug 18.1 & 0.6 & Swift & $2.41\times 10^{8}$ & $< 7.8 \times 10^{-3}$ & \nodata & N & \cite{evans2017} \\
    2017 Aug 18.2 & 0.7 & NuSTAR & $1.20\times 10^{9}$ & $< 7.3 \times 10^{-4}$ & \nodata & N & \cite{evans2017} \\
    2017 Aug 18.5 & 1.0  & Swift & $2.41\times 10^{8}$ & $< 7.5 \times 10^{-2}$ & \nodata & N & \cite{evans2017} \\
    2017 Aug 18.6 & 1.1 & Swift & $2.41\times 10^{8}$ & $< 5.0 \times 10^{-3}$ & \nodata & N & \cite{evans2017} \\
    2017 Aug 19.0 & 1.5 & Swift & $2.41\times 10^{8}$ & $< 3.7 \times 10^{-3}$ & \nodata & N & \cite{evans2017} \\
    2017 Aug 19.6 & 2.1 & Swift & $2.41\times 10^{8}$ & $< 2.9 \times 10^{-3}$ & \nodata & N & \cite{evans2017} \\
    2017 Aug 19.8 & 2.3 & Swift & $2.41\times 10^{8}$ & $< 3.8 \times 10^{-3}$ & \nodata & N & \cite{evans2017} \\
    2017 Aug 19.9 & 2.4 & Chandra &  $2.41\times10^{8}$ & $ <2.24\times10^{-4}$ & \nodata & N & \cite{margutti2017,troja2017,nynka2018}  \\
    2017 Aug 20.1 & 2.6 & Swift & $2.41\times 10^{8}$ & $< 4.0 \times 10^{-3}$ & \nodata & N & \cite{evans2017} \\
    2017 Aug 20.4 & 2.9 & Swift & $2.41\times 10^{8}$ & $< 1.1 \times 10^{-3}$ & \nodata & N & \cite{evans2017} \\
    2017 Aug 21.1 & 3.6 & Swift & $2.41\times 10^{8}$ & $< 1.9 \times 10^{-3}$ & \nodata & N & \cite{evans2017} \\
    2017 Aug 21.9 & 4.4 & NuSTAR & $1.20\times 10^{9}$ & $< 5.8 \times 10^{-4}$ & \nodata & N & \cite{evans2017} \\
    2017 Aug 22.0 & 4.5 & Swift & $2.41\times 10^{8}$ & $< 1.8 \times 10^{-3}$ & \nodata & N & \cite{evans2017} \\
    2017 Aug 23.3 & 5.8 & Swift & $2.41\times 10^{8}$ & $< 2.0 \times 10^{-3}$ & \nodata & N & \cite{evans2017} \\
    2017 Aug 24.0 & 6.5 & Swift & $2.41\times 10^{8}$ & $< 2.2 \times 10^{-3}$ & \nodata & N & \cite{evans2017} \\
    2017 Aug 26.7 & 9.2 & Chandra &  $2.41\times10^{8}$ &  $ 4.48\times10^{-4}$ & $^{+ 1.44}_{- 1.19}\times10^{-4}$ & Y & \cite{troja2017,margutti2017,nynka2018,hajela2019}\\
    2017 Aug 27.0 & 9.5 & Swift & $2.41\times 10^{8}$ & $< 2.5 \times 10^{-3}$ & \nodata & N & \cite{evans2017} \\
    2017 Aug 28.4 & 10.9 & Swift & $2.41\times 10^{8}$ & $< 4.0 \times 10^{-3}$ & \nodata & N & \cite{evans2017} \\
    2017 Aug 29.0 & 11.5 & Swift & $2.41\times 10^{8}$ & $< 1.7 \times 10^{-3}$ & \nodata & N & \cite{evans2017} \\
    2017 Aug 30.0 & 12.5 & Swift & $2.41\times 10^{8}$ & $< 1.6 \times 10^{-3}$ & \nodata & N & \cite{evans2017} \\
    2017 Aug 31.1 & 13.6 & Swift & $2.41\times 10^{8}$ & $< 1.1 \times 10^{-3}$ & \nodata & N & \cite{evans2017} \\
    2017 Sep 01.2 & 14.7 & Swift & $2.41\times 10^{8}$ & $< 1.3 \times 10^{-3}$ & \nodata & N & \cite{evans2017} \\
    2017 Sep 01.4 & 14.9 & Chandra &  $2.41\times10^{8}$ &  $ 5.11\times10^{-4}$ & $^{+ 1.02}_{- 0.90}\times10^{-4}$ & Y  & \cite{troja2017,margutti2017,haggard2017,nynka2018,hajela2019}\\
    2017 Sep 02.4 & 15.9 & Swift & $2.41\times 10^{8}$ & $< 4.3 \times 10^{-3}$ & \nodata & N & \cite{evans2017} \\
    2017 Sep 04.7 & 18.2 & NuSTAR & $1.20\times 10^{9}$ & $< 1.8 \times 10^{-3}$ & \nodata & N & \cite{evans2017} \\
    2017 Sep 05.6 & 19.1 & NuSTAR & $1.20\times 10^{9}$ & $< 1.2 \times 10^{-3}$ & \nodata & N & \cite{evans2017} \\
    2017 Sep 06.7 & 20.2 & NuSTAR & $1.20\times 10^{9}$ & $< 9.2 \times 10^{-4}$ & \nodata & N & \cite{evans2017} \\
    2017 Sep 21.5 & 35.0 & NuSTAR & $1.20\times 10^{9}$ & $< 4.6 \times 10^{-4}$ & \nodata & N & \cite{evans2017} \\
    2017 Nov 28 & 103 & NuSTAR &  $1\times10^{9}$ &  $<2.6\times 10^{-3}$ & \nodata & N & \cite{troja2018}\\
    2017 Nov 28 & 103 & NuSTAR &  $3\times10^{9}$ &  $<2.0\times 10^{-3}$ & \nodata & N & \cite{troja2018}\\
    2017 Dec 04 & 109 & Chandra &  $2.41\times10^{8}$ &  $21.17\times10^{-4}$ & $^{+ 1.90}_{- 1.80}\times10^{-4}$ & Y & \cite{ruan2018,margutti2018,troja2018,nynka2018}\\
    2017 Dec 29 & 133 & XMM-Newton &  $2.41\times10^{8}$ &  $21.94\times10^{-4}$ & $^{+ 4.42}_{- 4.15}\times10^{-4}$ & Y & \cite{davanzo2018}\\
    2018 Jan 23 & 158 & Chandra &  $2.41\times10^{8}$ &  $21.87\times10^{-4}$ & $^{+ 1.88}_{- 1.78}\times10^{-4}$ & Y & \cite{troja2018,margutti2018,nynka2018}\\
    2018 Jan 26 & 161 & XMM-Newton &  $2.41\times10^{8}$ &  $17.38\times10^{-4}$ & $^{+ 2.89}_{- 2.76}\times10^{-4}$ & Y & \cite{piro2019}\\
    2018 May 04 & 259 & Chandra &  $2.41\times10^{8}$ &  $11.50\times10^{-4}$ & $^{+ 1.45}_{- 1.34}\times10^{-4}$ & Y & \cite{nynka2018,piro2019,hajela2019}\\
    2018 Aug 10 & 357 & Chandra &  $2.41\times10^{8}$ &  $ 7.12\times10^{-4}$ & $^{+ 1.45}_{- 1.27}\times10^{-4}$ & Y & \cite{troja2019,hajela2019}\\
    2019 Mar 22 & 581 & Chandra &  $2.41\times10^{8}$ &  $ 2.63\times10^{-4}$ & $^{+ 0.77}_{- 0.64}\times10^{-4}$ & Y & \cite{hajela2019,troja2020-paper}\\
    2019 Aug 29 & 741 & Chandra &  $2.41\times10^{8}$ &  $ 2.19\times10^{-4}$ & $^{+ 0.81}_{- 0.65}\times10^{-4}$ & Y & \cite{hajela2019,troja2020-paper}\\
    2020 Mar 13 & 938 & Chandra &  $2.41\times10^{8}$ &  $ 1.30\times10^{-4}$ & $^{+ 0.59}_{- 0.46}\times10^{-4}$ & Y & \cite{hajela2020,troja2020-paper}\\
    \hline
    \hline
    \multicolumn{8}{c}{Optical (HST) measurements} \\ \hline
    2017 Dec 4   &  109  &  HST/F160W  &  1.88$\times10^{5}$ &  $<0.363$  &    \nodata      &  N & \cite{lyman2018} \\
    2017 Dec 4   &  109  &  HST/F814W  &  3.80$\times10^{5}$ &  0.109     &  0.017   & Y & \cite{lyman2018} \\  
    2017 Dec 4   &  109  &  HST/F140W  &  2.14$\times10^{5}$ &  $<0.276$  & \nodata & N  & \cite{lyman2018} \\  
    2017 Dec 6   &  111  &  HST/F606W  &  5.06$\times10^{5}$ &  0.111     &  0.019   &  N & \cite{fong2019}  \\  
    2018 Jan 2   &  137  &  HST/F606W  &  5.06$\times10^{5}$ &  0.084     &  0.018   &  N & \cite{fong2019}  \\  
    2018 Jan 29  &  165  &  HST/F606W  &  5.06$\times10^{5}$ &  0.091     &   0.016  &  N & \cite{fong2019,piro2019}  \\
    2018 Feb 5   &  170  &  HST/F814W  &  3.80$\times10^{5}$ &  0.113     &  0.019   & Y & \cite{lamb2019}  \\  
    2018 Feb 5   &  172  &  HST/F606W  &  5.06$\times10^{5}$ &  0.085     &  0.017   & N  & \cite{fong2019,lamb2019}  \\
    2018 Mar 14  &  209  &  HST/F606W  &  5.06$\times10^{5}$ &  0.082     &  0.020   & N  & \cite{fong2019,piro2019}  \\
    2018 Mar 24  &  218  &  HST/F606W  &  5.06$\times10^{5}$ &  0.063     &  0.018   & N  & \cite{fong2019}  \\  
    2018 Jun 10  &  297  &  HST/F606W  &  5.06$\times10^{5}$ &  0.044     &  0.014   & N  & \cite{fong2019,lamb2019}  \\  
    2018 Jul 11  &  328  &  HST/F606W  &  5.06$\times10^{5}$ &  0.034     &  0.011   & N  & \cite{fong2019,lamb2019}  \\
    2018 Jul 20  &  337  &  HST/F606W  &  5.06$\times10^{5}$ &  $<0.048$  & \nodata & N  & \cite{fong2019}  \\
    2018 Aug 08  &  355  &  HST/F814W  &  3.80$\times10^{5}$ &  $<0.058$  & \nodata & \nodata  & This work (PI: Tanvir) \\
    2018 Aug 15  &  362  &  HST/F606W  &  5.06$\times10^{5}$ &  0.027     &  0.007   & N  & \cite{fong2019,lamb2019}  \\
    2019 Mar 24  &  584  &  HST/F606W  &  5.06$\times10^{5}$ &  $<0.019$  & \nodata  & N  & \cite{fong2019} \\
    \hline
\multicolumn{8}{l}{Notes: Correction factors of 0.6 and 0.8 have been applied to the eMERLIN and ATCA flux density values respectively.} 
\end{longtable*}
\end{center}

\begin{table*}
\centering
\caption{ \chandra\ and \xmm\ observational data}
\label{tab:table_xray_obsdata}
\begin{center}
\begin{tabular}{c c c c c c}
\hline
ObsID	& Exposure  & Start date & PI & Count rate & Flux \\
        & (ks)      &       (UT)      &    & (ks$^{-1}$) & (10$^{-14}$ \ergcms) \\
\hline
\multicolumn{6}{c}{\chandra\ observations}\\
\hline
18955 & 24.64 &      19-Aug-   2017  &  Fong     &  $<0.12$  &  $<0.27$\\
19294 & 49.41 &      26-Aug-   2017 &  Troja    &  0.24$\pm$ 0.07 &  0.52$^{+ 0.29}_{- 0.21}$\\
20728 & 46.69 &       1-Sep-   2017 &  Troja    &  0.34$\pm$ 0.09 &  0.66$^{+ 0.31}_{- 0.24}$\\
18988 & 46.69 &       2-Sep-   2017 &  Haggard  &  0.25$\pm$ 0.07 &  0.51$^{+ 0.29}_{- 0.21}$\\
20860 & 74.09 &       3-Dec-   2017 &  Wilkes   &  1.34$\pm$ 0.13 &  2.55$^{+ 0.44}_{- 0.40}$\\
20861 & 24.74 &       6-Dec-   2017 &  Wilkes   &  1.25$\pm$ 0.22 &  2.44$^{+ 0.80}_{- 0.65}$\\
20936 & 31.75 &      17-Jan-   2018 &  Wilkes   &  1.63$\pm$ 0.23 &  3.46$^{+ 0.85}_{- 0.73}$\\
20938 & 15.86 &      21-Jan-   2018 &  Wilkes   &  1.70$\pm$ 0.33 &  3.23$^{+ 1.14}_{- 0.92}$\\
20937 & 20.77 &      23-Jan-   2018 &  Wilkes   &  1.30$\pm$ 0.25 &  2.63$^{+ 0.92}_{- 0.75}$\\
20939 & 22.25 &      24-Jan-   2018 &  Wilkes   &  0.94$\pm$ 0.21 &  1.91$^{+ 0.77}_{- 0.61}$\\
20945 & 14.22 &      28-Jan-   2018 &  Wilkes   &  0.84$\pm$ 0.24 &  2.01$^{+ 1.12}_{- 0.81}$\\
21080 & 50.79 &       3-May-   2018 &  Wilkes   &  0.65$\pm$ 0.11 &  1.27$^{+ 0.40}_{- 0.33}$\\
21090 & 46 &       5-May-   2018 &  Wilkes   &  0.75$\pm$ 0.13 &  1.43$^{+ 0.44}_{- 0.37}$\\
21371 & 67.17 &      10-Aug-   2018 &  Troja    &  0.41$\pm$ 0.08 &  0.83$^{+ 0.29}_{- 0.24}$\\
21322 & 35.64 &      21-Mar-   2019 &  Margutti  &  0.14$\pm$ 0.06 &  0.43$^{+ 0.41}_{- 0.25}$\\
22157 & 38.19 &      22-Mar-   2019 &  Margutti  &  0.15$\pm$ 0.06 &  0.31$^{+ 0.27}_{- 0.17}$\\
22158 & 24.93 &      23-Mar-   2019 &  Margutti  &  0.08$\pm$ 0.06 &  0.31$^{+ 0.54}_{- 0.25}$\\
21372 & 40 &      27-Aug-   2019 &  Troja    &  0.02$\pm$ 0.03 &  0.06$^{+ 0.08}_{- 0.08}$\\
22736 & 33.61 &      29-Aug-   2019 &  Troja    &  0.09$\pm$ 0.05 &  0.27$^{+ 0.35}_{- 0.19}$\\
22737 & 25.25 &      30-Aug-   2019 &  Troja    &  0.16$\pm$ 0.08 &  0.88$^{+ 0.95}_{- 0.55}$\\
21323 & 24.29 &       9-Mar-   2020 &  Margutti  &  0.08$\pm$ 0.06 &  0.34$^{+ 0.58}_{- 0.27}$\\
23183 & 16.28 &   13-Mar-   2020  &  Margutti  &  $<0.18$  &  $<0.57$\\
23184 & 19.85 &      15-Mar-   2020 &  Margutti  &  0.10$\pm$ 0.07 &  0.64$^{+ 1.08}_{- 0.50}$\\
23185 & 36.18 &      15-Mar-   2020 &  Margutti  &  0.02$\pm$ 0.03 &  0.07$^{+ 0.09}_{- 0.09}$\\
\hline
\multicolumn{6}{c}{\xmm\ observations}\\
\hline
811210101 & 26.36 &      29-Dec-   2017 & Schartel &  2.17$\pm$ 0.40 &  2.56$^{+ 0.85}_{- 0.78}$\\
811212701 & 48.12 &      26-Jan-   2018 & Schartel &  1.82$\pm$ 0.27 &  2.03$^{+ 0.57}_{- 0.52}$\\
\hline
\end{tabular}
\end{center}
\tablecomments{Count rates are observed (not corrected for PSF losses) in the 0.5--8 keV band for \chandra\ and 0.2--10 keV band for \xmm. Fluxes are in the 0.3--10 keV band and are corrected for Galactic absorption.}
\end{table*}

\begin{table}[]
\centering
\begin{tabular}{lll}
\hline \hline
Parameter      & Value & Units\\
\hline
\multicolumn{3}{c}{Broken power-law / Analytical jet model}\\
\hline
$F_{\nu,p}$ & $101\pm3$ & $\mu$Jy \\
$t_p$       & $155\pm4$ & days \\
$\alpha_1$  & $0.86\pm0.04$  \\
$\alpha_2$  & $-1.92^{+0.10}_{-0.12}$ \\
log$_{10}(s)$ & $0.56^{+0.12}_{-0.11}$ \\
$\beta$     & $-0.584\pm0.002$ \\
$\chi^2$/dof   & 75/97 \\
p              & $2.168\pm0.004$\\
$E/n_{\rm ISM}$ & $\simeq 1.5 \times 10^{53}$ & erg\,cm$^3$\\
$\theta_{\rm v}$  & $\simeq14-20$ & degrees\\
$\theta_{\rm j}$  & $\simeq1-4$   & degrees\\
\hline
\multicolumn{3}{c}{Structured jet model}\\
\hline
$\theta_{\rm v}$  &  $35.2\pm0.6$ & degrees\\
$\epsilon_e$   &  $7.8_{-0.6}^{+1.0}\times10^{-3}$\\
$\epsilon_b$   &   $9.9_{-2.2}^{+4.7}\times10^{-4}$\\
$p$              & $2.07_{-0.02}^{+0.01}$\\
$n_{\rm ISM}$  & $9.8_{-1.6}^{+0.2}\times10^{-3}$ & cm$^{-3}$ \\
$\chi^2$/dof   & 95.2/97 \\
\hline
\end{tabular}
\caption{Parameters estimated from modeling the afterglow light curve.}
\label{tab:best-fit}
\end{table}

\begin{table*}
    \centering
    \caption{Summary of published non-thermal afterglow modeling for \object{GW170817}}
    \begin{tabular}{ccccccccc}
    \hline
    \hline
       Reference  & Model & $\theta_{v}$ &  $ \theta_j$ & $\log_{10}E_{0,52}$ & $\log_{10}n_0$ & $log_{10}\epsilon_e$ & $log_{10}\epsilon_B$ & VLBI fit?\\
    & & (deg) & (deg) & (erg) & (cm$^{-3}$) & \\
    \hline
 This work & SJ & $35.2_{+0.6}^{-0.6}$ & \nodata & $-0.2$ & $-2.0^{+0.1}_{-0.1}$ & $-1.1^{+0.1}_{-0.1}$ & $-3.0^{+0.1}_{-0.2}$ & N\\
 \cite{lazzati2018} & SJ & $33.0^{+4.0}_{-2.5}$ & \nodata & \nodata & $-2.4^{+2.1}_{-2.8}$ & $-1.2^{+1.0}_{-1.0}$ & $-2.5^{+0.7}_{-0.7}$ & N\\
\cite{hajela2019} & BF & $30.4^{+4.0}_{-3.4}$ & $5.9^{+1.0}_{-0.7}$ & $-1.3^{+0.0}_{-1.0}$ & $-2.6^{+0.4}_{-0.6}$ & $-0.8^{+0.4}_{-0.6}$ & $-2.6^{+0.9}_{-1.2}$ & N\\
\cite{wu2019} & BF & $30.3^{+0.7}_{-0.4}$ & $6.3^{+1.1}_{-1.7}$ & ${-0.8}^{+0.2}_{-1.4}$ & $-2.0^{+0.7}_{-1.0}$ & $-1.0^{+0.6}_{-0.9}$ & $-3.6^{+1.3}_{-1.4}$ & N\\
\cite{resmi2018} & GJ & $26.9^{+8.6}_{-4.6}$ & $6.9^{+2.3}_{-1.7}$ & $-0.2^{+0.4}_{-0.5}$ & $-2.7^{+0.9}_{-1.0}$ & $-0.6^{+0.1}_{-0.5}$ & $-4.4^{+1.1}_{-0.5}$ & N\\
\cite{ryan2020} & PLJ & $25.2^{+6.9}_{-6.9}$ & $2.6^{+0.7}_{-0.7}$ & $0.9^{+1.1}_{-0.8}$ & $-2.6^{+1.1}_{-1.1}$ & $-1.2^{+0.7}_{-1.2}$ & $-3.8^{+1.1}_{-0.9}$ & N\\
" & GJ & $22.9^{+6.3}_{-6.3}$ & $3.8^{+1.0}_{-1.0}$ & $1.0^{+1.0}_{-0.7}$ &  $-2.7^{+1.0}_{-1.0}$ & $-1.4^{+0.7}_{-1.1}$ & $-4.0^{+1.1}_{-0.7}$ & N\\
\cite{troja2019} & GJ & $21.8^{+6.3}_{-6.3}$ & $3.4^{+1.0}_{-0.6}$ & $0.8^{+0.9}_{-0.6}$ & $-2.5^{+0.9}_{-1.0}$ & $-1.4^{+0.5}_{-0.6}$ & $-4.0^{+1.0}_{-0.7}$ & N\\
\cite{lamb2019} & 2C & $20.6^{+1.7}_{-1.7}$ & $4.0^{+0.6}_{-0.6}$ & $0.0^{+0.9}_{-0.6}$ & $-3.3^{+0.6}_{-1.6}$ & $-1.3^{+0.6}_{-0.7}$ & $-2.4^{+1.4}_{-0.9}$ & N\\
 " & GJ & $19.5^{+1.1}_{-1.1}$ & $5.2^{+0.6}_{-0.6}$ &  $0.4^{+0.5}_{-0.4}$ & $-4.1^{+0.5}_{-0.5}$ & $-1.4^{+0.5}_{-0.6}$ & $-2.1^{+0.8}_{-1.0}$ & N\\
\cite{hotokezaka2019} & GJ & $16.6^{+0.6}_{-0.6}$ & $3.4^{+0.6}_{-0.6}$ & \nodata & \nodata & \nodata & \nodata & Y\\
 " & PLJ & $16.6^{+0.6}_{-0.6}$ & $2.9^{+0.6}_{-0.6}$ & \nodata & \nodata & \nodata & \nodata & Y\\
\cite{ghirlanda2019} & PLJ & $15.0^{+1.5}_{-1.0}$ & $3.4^{+1.0}_{-1.0}$ & $0.4^{+0.6}_{-7.0}$ & $-3.6^{+0.7}_{-0.7}$ & \nodata & $-3.9^{+1.6}_{-1.6}$ & Y\\
  \cite{mooley2018-vlbi} & HD & [$14,28$] & $<$5 & [$0,1.5$] & [$-4,-2.3$] & 0.1 & [$-2,-5$] & Y\\
\hline
\multicolumn{9}{l}{\scriptsize GJ=Gaussian Jet, PLJ=Power-law jet, SJ=Other structured jets, BF=Boosted fireball, HD=Hydrodynamic simulations, 2C=Two component}
    \end{tabular}
    \label{tab:modeling_summary}
\end{table*}
\newpage
\bibliography{ref}

\end{document}